\documentclass[preprint,5p,times,twocolumn]{elsarticle}

\usepackage[utf8]{inputenc}
\usepackage{amsmath}
\usepackage{amsfonts}
\usepackage{mathrsfs}
\usepackage{amssymb}
\usepackage{xfrac}
\usepackage{graphicx}
\usepackage{enumitem}
\usepackage{booktabs}
\usepackage{hyperref}
\usepackage{setspace}
\usepackage{multirow}
\usepackage{lineno}
\usepackage{array}

\usepackage{xcolor}
\hypersetup{
	colorlinks,
	linkcolor={black},
	citecolor={blue!50!black},
	urlcolor={blue!80!black}
}

\newcolumntype{C}[1]{>{\centering\arraybackslash}m{#1}}
\newcolumntype{L}[1]{>{\arraybackslash}m{#1}}

\AtBeginDocument{
	\heavyrulewidth=.08em
	\lightrulewidth=.05em
	\cmidrulewidth=.03em
	\belowrulesep=.65ex
	\belowbottomsep=0pt
	\aboverulesep=.4ex
	\abovetopsep=0pt
	\cmidrulesep=\doublerulesep
	\cmidrulekern=.5em
	\defaultaddspace=.5em
}

\DeclareMathOperator\arctanh{arctanh}
\newcommand{\degrees}{\ensuremath{^\circ}}

\makeatletter
\def\@author#1{\g@addto@macro\elsauthors{\normalsize%
    \def\baselinestretch{1}%
    \upshape\authorsep#1\unskip\textsuperscript{%
      \ifx\@fnmark\@empty\else\unskip\sep\@fnmark\let\sep=,\fi
      \ifx\@corref\@empty\else\unskip\sep\@corref\let\sep=,\fi
    }%
    \def\authorsep{\unskip,\space}%
    \global\let\@fnmark\@empty
    \global\let\@corref\@empty  
  \global\let\sep\@empty}%
  \@eadauthor={#1}
}
\makeatother

\journal{Nuclear Instruments and Methods in Research A}

\begin{document}
	
\begin{frontmatter}
  \author[JLab]{J.D.~Maxwell\corref{cor1}}
\ead{jmaxwell@jlab.org}
\author[TU,ANL]{W.R.~Armstrong}
\author[SNL]{S.~Choi}
\author[JLab]{M.K.~Jones}
\author[SNL]{H.~Kang}
\author[HU]{A.~Liyanage}
\author[TU]{Z.-E.~Meziani}
\author[UVa]{J.~Mulholland}
\author[MSU]{L.~Ndukum}
\author[UVa]{O.A.~Rond\'{o}n}
% ...

\address[JLab]{Thomas Jefferson National Accelerator Facility, Newport News, VA}
\address[TU]{Temple University, Philadelphia, PA}
\address[UVa]{University of Virginia, Charlottesville, VA}
% ...
\address[SNL]{Seoul National University, Seoul, Korea}
\address[HU]{Hampton University, Hampton, VA}
\address[MSU]{Mississippi State University, Starkville, MS}
% ...
\address[ANL]{Argonne National Laboratory, Argonne, IL}
% ...

\cortext[cor1]{Corresponding author}
\cortext[dec]{Deceased}

  \title{Design and Performance of the Spin Asymmetries of the Nucleon Experiment} 		

  \begin{abstract}
    The Spin Asymmetries of the Nucleon Experiment (SANE) performed inclusive, 
    double-polarized electron scattering measurements of the proton at the 
    Continuous Electron Beam Accelerator Facility at Jefferson Lab. A novel detector array 
    observed scattered electrons of four-momentum transfer $2.5 < Q^2< 
    6.5$\,GeV$^2$ and Bjorken scaling $0.3<x<0.8$ from initial beam energies of 4.7 
    and 5.9\,GeV.  Employing a polarized proton target whose magnetic field direction could be rotated with 
    respect to the incident electron beam, both parallel and near perpendicular 
    spin asymmetries were measured, allowing model-independent access to transverse 
    polarization observables $A_1$, $A_2$, $g_1$, $g_2$ and moment $d_2$ of the 
    proton. This document summarizes the operation and performance of the polarized 
    target, polarized electron beam, and novel detector systems used during the 
    course of the experiment, and describes analysis techniques utilized to access 
  the physics observables of interest.  \end{abstract}


\end{frontmatter}


\section{Introduction}

Deep-inelastic leptonic scattering has driven the study of nucleon spin 
structure as the cleanest probe available to hadronic physics. Inclusive spin 
asymmetry measurements at high $x$ offer a particularly clear view of nucleon 
structure where the influence of sea quarks falls away. The Spin 
Asymmetries of the Nucleon Experiment (SANE) was devised to precisely measure 
inclusive double-spin asymmetries $A_1^p$ and $A_2^p$ in the deep-inelastic region of final state invariant mass $W$ and 
in a wide range of $x$, allowing direct access to spin structure functions 
$g_1^p$ and the higher-twist dependent $g_2^p$, revealing trends as $x$ 
approaches unity, and connecting spin structure function moments to lattice QCD 
calculations. Where a thorough exploration of these asymmetries with 
traditional, narrow-acceptance spectrometer techniques would be a protracted, 
expensive effort, SANE viewed a wide kinematic range using a novel, 
non-magnetic, high-acceptance electron detector array. 
This array utilized the drift space between a Cherenkov detector and an
electromagnetic calorimeter to create a ``telescope'' to isolate
electron events produced in the target from possible background produced elsewhere
 along the beamline. To access both spin asymmetries in a model independent way, 
a polarized proton target was needed which could provide both longitudinal and 
the more challenging transverse target orientation components.


SANE was performed in Hall C of the Thomas Jefferson National Accelerator Facility from 
January to March of 2009. A polarized electron beam at energies of 4.7 or 5.9 
GeV was incident on a solid, polarized proton target to produce spin 
asymmetries with the target polarized parallel to the beam, or nearly 
perpendicular (80\degrees) to it. Scattered electrons were observed using Hall 
C's standard High Momentum Spectrometer (HMS), as well as a novel detector 
system, the Big Electron Telescope Array (BETA), resulting in a kinematic 
coverage of $2.5 < Q^2< 6.5$\,GeV$^2$ and $0.3<x<0.8$. While BETA was built 
with SANE's primary aim in mind---accessing deep-inelastic double spin 
asymmetries---the HMS also allowed two additional, single-arm measurements to 
be performed opportunistically during the experiment. Measurements of spin 
asymmetries $A_1^p$ and $A_2^p$ were performed by the HMS in the resonance and 
low-$W$ DIS regions, and the ratio of the electric to magnetic proton elastic form factors was measured using 
HMS--BETA coincidences as well as HMS single-arm data.

This document describes the design of SANE, with emphasis on its non-standard 
additions to Jefferson Lab's Hall C, as well as the performance of each system 
during the experiment. We also give an overview of the analysis and corrections 
needed to produce spin asymmetries from BETA. 

\section{Polarized Electron Beam}

Jefferson Lab's Continuous Electron Beam Accelerator Facility (CEBAF) consists 
of two linear accelerators, which at the time of this experiment, each 
accelerated electrons by roughly 600\,MeV. Recirculating arcs connect these 
linacs, allowing a nominal 6\,GeV maximum beam energy after 5 passes around the 
``race-track''~\cite{Leeman}. Laser-excited, strained GaAs photocathodes provided a polarized 
electron source which switched helicity in 30\,Hz pseudo-random batches.  The beam current delivered to Hall C was 
limited to below 100\,nA by the heat and radiation dose generated in the solid polarized target.

\subsection{Hall C Beamline}

\begin{figure}	
  \begin{center}
    \includegraphics[width=0.48\columnwidth]{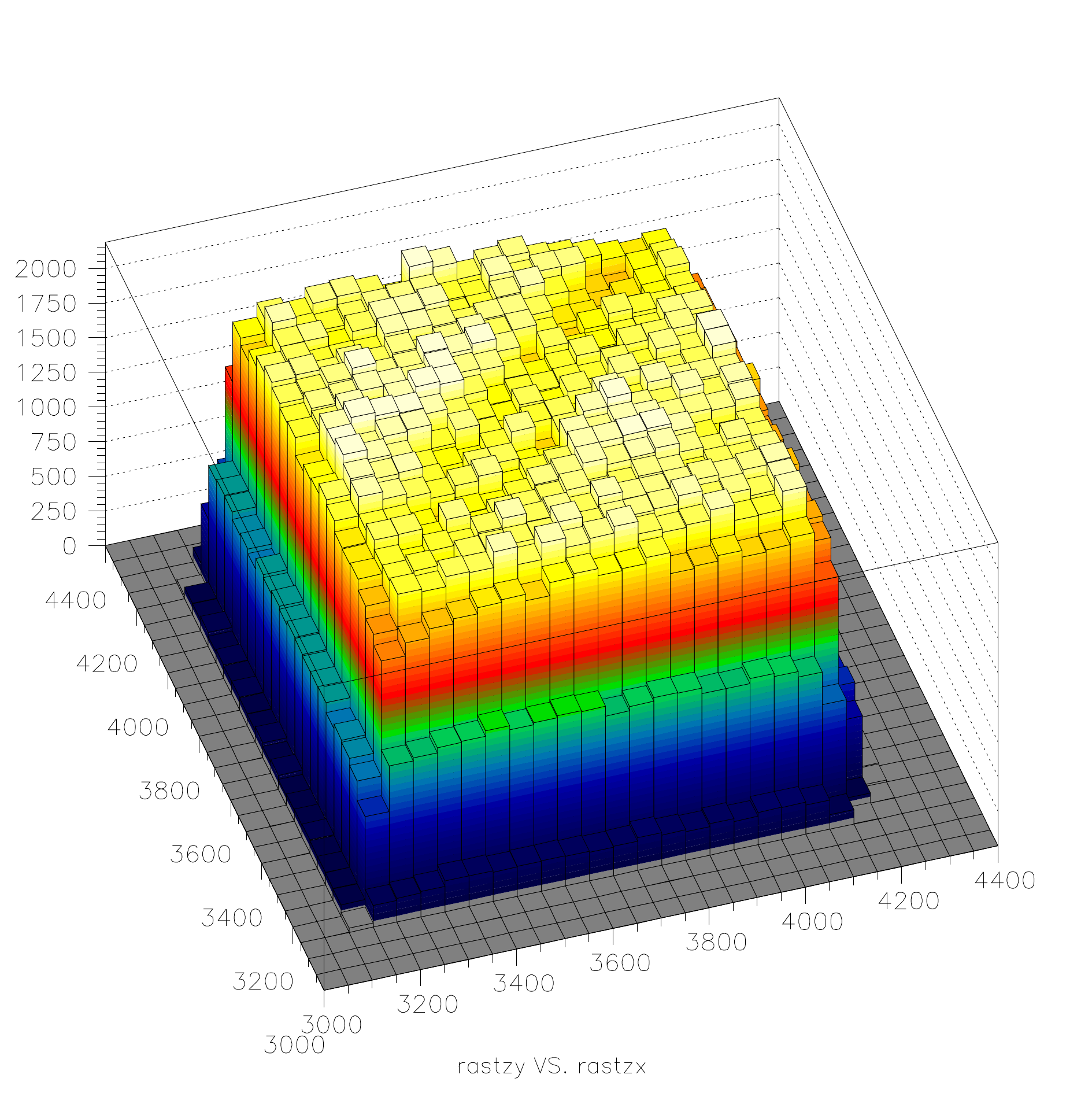}
    \includegraphics[width=0.48\columnwidth]{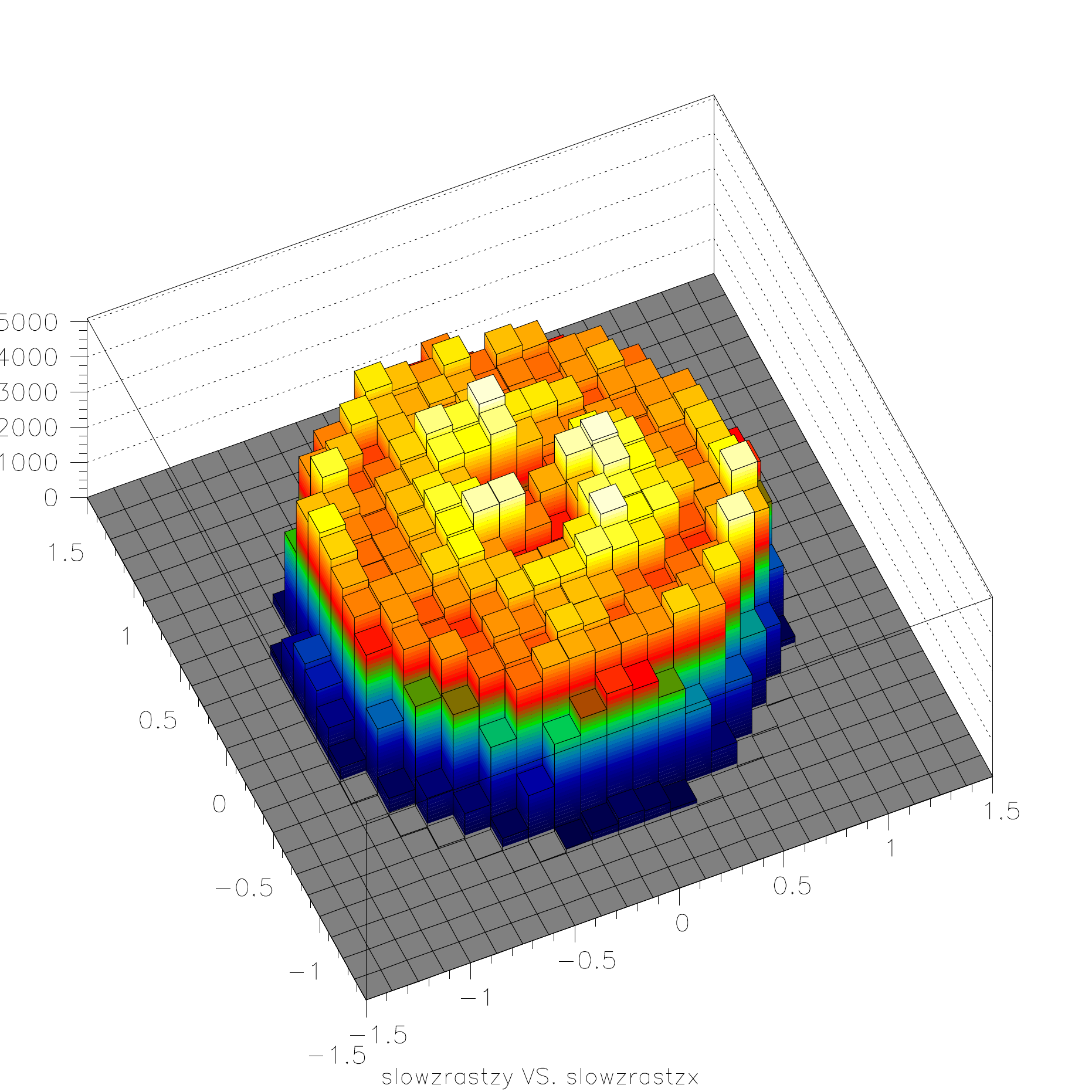}
  \end{center}
  \caption{Magnitude of hits the detector system versus the ``fast'' (left) and ``slow'' (right) raster positions, showing the raster patterns for a typical run. At left, $x$ and $y$ are given in ADC channels, where 500 channels = 1\,mm; at right, $x$ and $y$ units are in cm. }
  \label{fig:raster}
\end{figure}

Upon entering Hall C, the beam was expanded from below 100\,$\mu$m in diameter to 
a $2\times2$\,mm$^2$ square by two air-core magnets roughly 25\,m upstream of the 
target, producing the ``fast raster''~\cite{Yan199546}. To further retard 
damage to the target polarization by radiation from the beam, an additional, 
circular ``slow raster'' was created by scanning the beam over a 2.0\,cm  
diameter spiral pattern to better cover the 2.5\,cm diameter target cell~\cite{Fukuda199745}.  
Figure~\ref{fig:raster} shows each raster pattern as observed from hits in the 
BETA detector versus the recorded raster amplitude.

To counteract the bending of the beam down and away as it approached the target center while
 under the influence of the near perpendicular, 5\,T magnetic field, it was passed through two dipole chicane 
magnets, BE and BZ, which bent the beam down and then up towards the scattering 
chamber, respectively.  Table~\ref{tab:chicane} shows the  deflection of the 
two chicane magnets for both energy settings used while the target was in its 
near perpendicular configuration. Any out of plane precession of the electron 
spins due to the chicane transport is canceled as the beam is subsequently bent in the opposite
 sense by the target magnet, so the beam polarization remains unaffected.

\begin{table}[ht]
  \begin{center}
    \begin{tabular}{ccccccc}
      \toprule

      Beam $E$ & BE Bend& BZ  Bend & Target Bend   \\ \midrule 
      4.7 GeV& -0.878$\degrees$ & 3.637$\degrees$ & -2.759$\degrees$  \\ 
      5.9 GeV& -0.704$\degrees$ & 2.918$\degrees$ & -2.214$\degrees$   \\
      \bottomrule
    \end{tabular}
    \caption[Table of chicane parameters.]{Table of chicane 
      parameters for 80$\degrees$ field for both beam energy 
    settings. Negative angles indicate downward bends. The target bending angle listed is that during the approach of the beam, not the bend after the beam passes through the target center.}
    \label{tab:chicane}
  \end{center}
\end{table}

After passing through the target, the electron beam was again deflected downwards. Rather than using a second set of chicane magnets to direct the beam up 
to the beam dump, an 80-foot long helium bag was devised to transport the beam 
to a temporary beam dump on the experimental floor. 

\subsection{Beam Polarization Measurement}

The beam polarization direction as it arrived in Hall C was not always 100\% longitudinal
due to the requirement to share polarization with the other experimental halls. The
degree of longitudinal polarization was a function of both the polarization direction
as the electrons left the injector, as set with a Wien filter, and the amount of spin precession through the accelerator before arrival in Hall C. The
precession itself is a function of the number of passes through the accelerator, the
overall beam energy, and the difference in energy between the two linear accelerators in the
machine.

The beam polarization was monitored in nine dedicated  M{\o}ller polarimeter 
measurements \cite{Hauger} covering each nominal beam energy and polarization 
setting. Periods of beam energy instability
during this experiment meant that the degree of spin precession through the machine was not
constant at a given energy setting, yielding more variation in the beam polarization with time
than is typically expected. Therefore, the nine polarization measurements were used to
interpolate the beam polarization throughout the experiment via a fit with three degrees of freedom:
the intrinsic polarization of the beam at the source $P_{\textrm{source}}$, the energy imbalance
of the north and south linear accelerators, and a small global correction to the overall
beam energy $F_{\textrm{corr}}$.  In addition, the beam polarization had been found to depend to some
degree on the quantum efficiency of the photocathode, which can be described by a
correction, $F(\epsilon_q)$, based on fits to data from the preceding experiment,  GEp-III \cite{PhysRevLett.104.242301}.
The beam polarization in Hall C, $P_B$, could then be expressed as a function of the Wien angle $\theta_w$,
quantum efficiency of the photocathode, and half wave plate status $n_{\textrm{hwp}}$, as
\begin{linenomath} \begin{equation}
P_B = (-1)^{n_{\textrm{hwp}}} P_{\textrm{source}} F_{\textrm{corr}} 
F(\epsilon_q) \cos (\theta_w + \varphi_{\textrm{precession}}),
\end{equation}\end{linenomath}
where $\varphi_{\textrm{precession}}$ is determined by following the spin 
precession through each bend in the accelerator. 

\begin{figure}	
  \begin{center}
    \includegraphics[width=0.9\columnwidth]{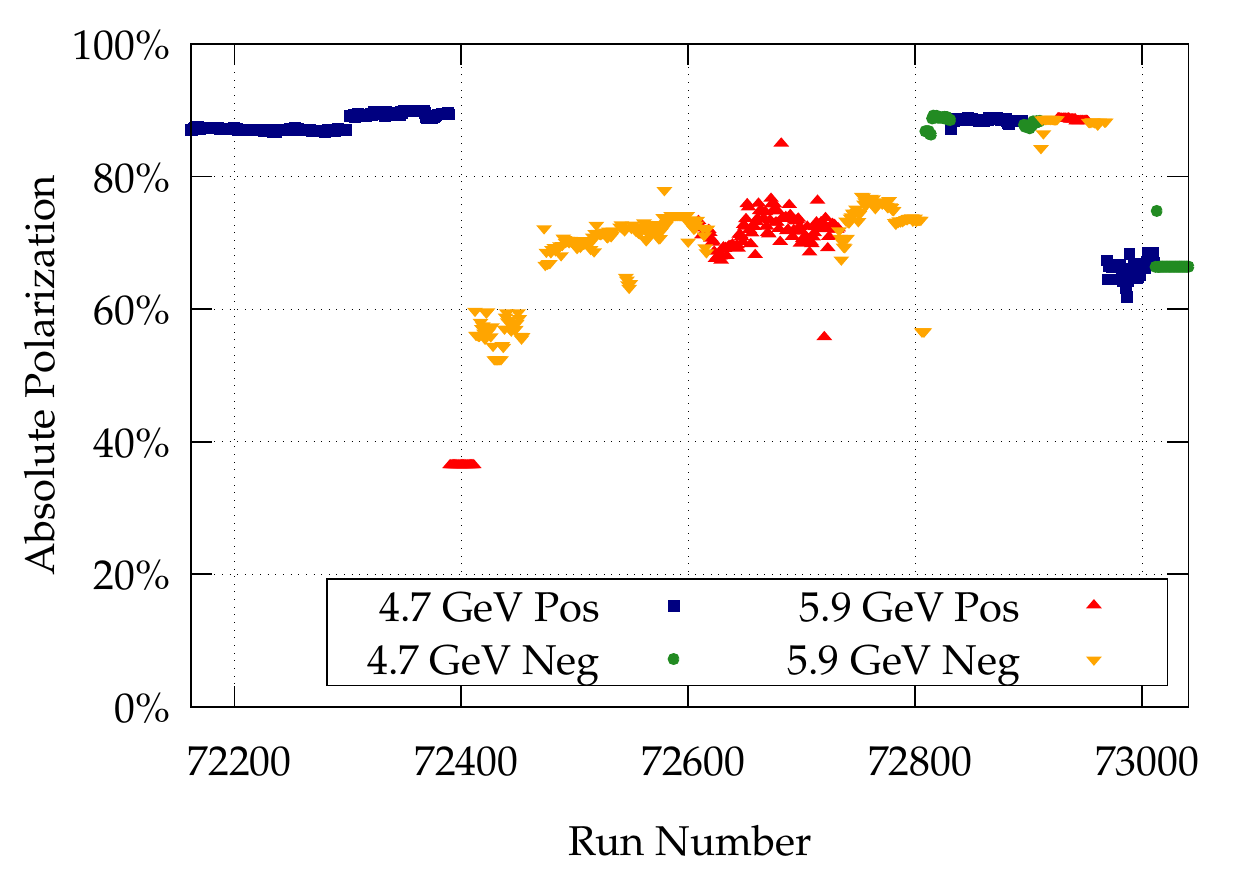}
  \end{center}
  \caption{Electron beam polarization per data-taking run.}
  \label{fig:beam_pol}
\end{figure}

Using the Wien angle, beam energy, quantum efficiency and half wave
plate status recorded over the course of each data-taking run,
the beam polarization over time was calculated using this fit.
By averaging these data over the charge accumulated on the target from beam current measurements at each moment in time, a
charge-averaged beam polarization was then produced for each
experimental run. For each beam energy, the Wien angle setting was chosen to maximum the combined figure
of merit for polarized beam to all JLab experimental halls. At beam energy of 4.7~GeV, the
Wien angle was set so that $P_B \approx P_{\textrm{source}}$ for Hall C and $P_B$ was not sensitive
to small changes in the beam energy.
Of note is
the rather low beam polarization near run 72400 at the beginning of the 5.9 GeV data taking, which came
from non-optimal setting of the Wien filter at the injector. The
increase in polarization that follows results from optimizing the
Wien angle. At 5.9 GeV,  the Wien angle was eventually optimized so $P_B \approx 0.8*P_{\textrm{source}}$, but
the $P_B$ had a small sensitivity to small changes in the beam energy which lead to
the fluctuations seen in Figure~\ref{fig:beam_pol}.


\section{Polarized Proton Target}

\begin{figure}		
	\begin{center}
		\includegraphics[width=0.8\columnwidth]{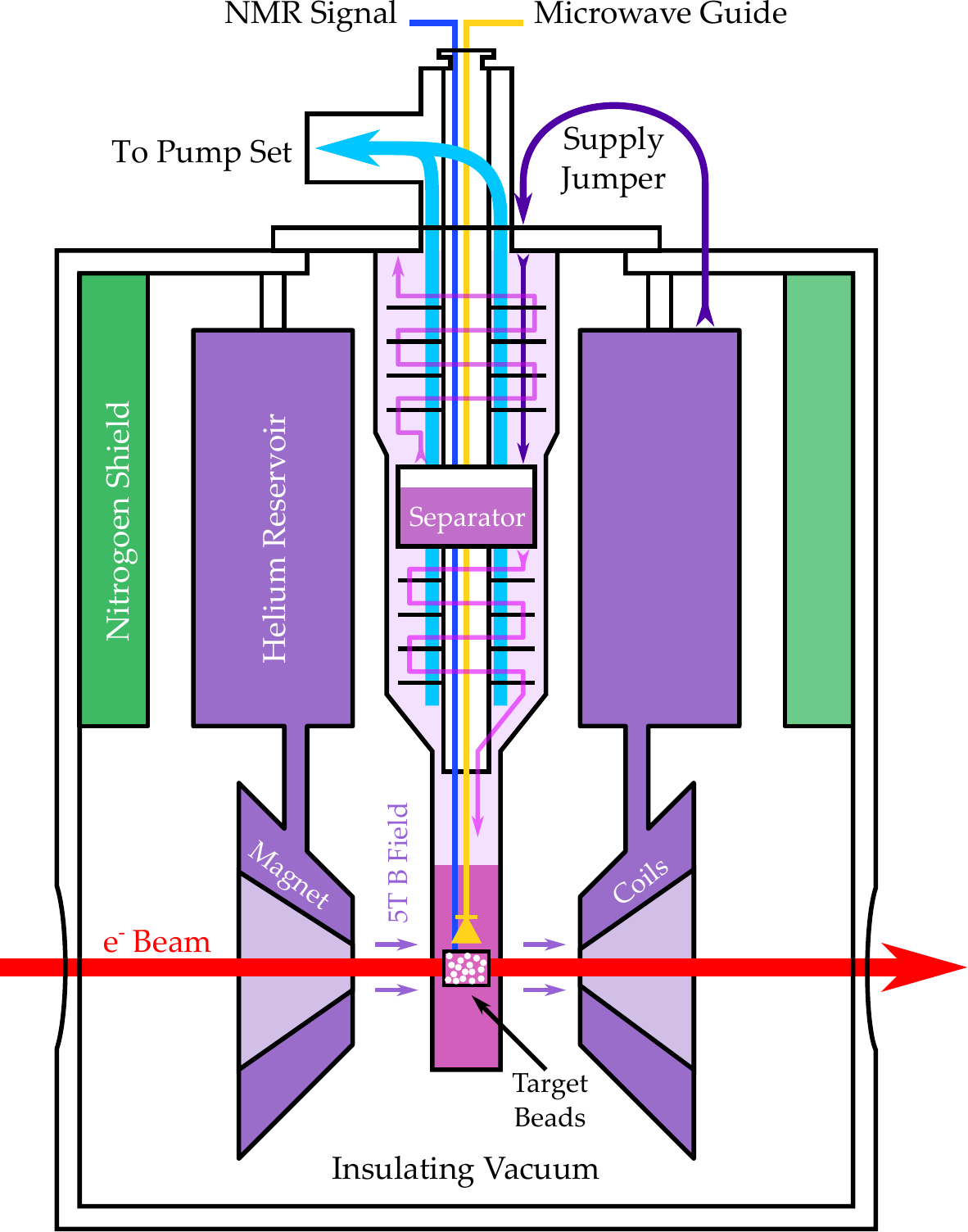}
	\end{center}
	\caption{Cross-sectional diagram of UVa polarized target cryostat, refrigerator, and 
		scattering chamber.}
	\label{fig:uvatar}
\end{figure}

SANE utilized the University of Virginia polarized solid target, which has had 
extensive use in electron scattering experiments at 
SLAC~\cite{CRABB19959,e155,e155x} and Jefferson 
Lab~\cite{PhysRevLett.87.081801,RSS,Pierce}, and is diagrammed in Figure~\ref{fig:uvatar}. Polarized protons were provided in 
the form of solid ammonia  (NH$_3$) beads held in one of two 2.5\,cm diameter, 2.5\,cm long 
cells (\textit{top} or \textit{bottom}) held in the ``nose'' of a helium 
evaporation refrigerator providing roughly 1\,W of cooling power at 1\,K. This 
nose was located at the center of an Oxford Instruments NbTi, 5\,T 
superconducting split pair magnet, which allowed beam passage parallel or 
perpendicular to the field. This magnet provided better than $10^{-4}$ field 
uniformity in the 3$\times$3$\times$3 cm$^3$ volume of the target scattering 
chamber. While the magnet allowed beam passage perpendicular to the field, 
the geometry of the coils did occlude the acceptance of BETA when oriented at 
90\degrees, so in practice 80\degrees\ was used. The field's alignment in Hall C
to its nominal values were to within 0.1 degree.

Polarized target nuclei were provided via dynamic nuclear polarization (DNP) of 
ammonia ($^{14}$NH$_3$). DNP employs high magnetic fields ($B\approx$ 5\,T) and 
low temperature ($T\approx$ 1\,K) to align spins in a target medium, using 
microwave radiation to drive polarizing transitions of coupled 
electron--nucleus spin states~\cite{crabb97}. These techniques offer excellent 
polarization of protons---exceeding 95\%---in a dense solid and can maintain 
this polarization under significant flux of ionizing radiation, such as an 
electron beam. 

At magnetic field $B$ and temperature $T$, the polarization of an ensemble of 
spin \sfrac{1}{2} particles is calculable by Boltzmann statistics as $P = \tanh (\mu B / (kT))$. At 
5\,T and 1\,K, this creates a high polarization of electron spins (99.8\%), but 
quite low polarization in protons (0.5\%). In DNP, microwave energy is 
used to transfer this high electron polarization to the proton spin system, which is accomplished
via several mechanisms, the simplest of which to explain is the solid-state effect~\cite{Abragam,Maly}. 
By taking advantage of coupling between free 
electron and proton spins, microwave radiation of frequency lower or higher 
than the electron paramagnetic resonance by the  proton magnetic resonance 
($\nu_{\textrm{EPR}}\pm\nu_{\textrm{NMR}}$) drives flip-flop transitions 
($e_\downarrow p_\downarrow \rightarrow e_\uparrow p_\uparrow$) to align or 
anti-align the proton with the field. The electron's millisecond relaxation 
time at 1\,K means that the free electron will relax quickly to become 
available to perform a polarizing flip-flop with another proton. While the 
protons take minutes to relax, they will frequently perform energy-conserving 
spin flip transitions via dipole--dipole coupling with other neighboring 
protons. This allows the transport of nuclear polarization away from the free 
electron sites---a process called ``spin-diffusion'' which tends to equalize 
the polarization throughout a material~\cite{Borghini}.

\subsection{Target polarization measurement}

The proton polarization was measured via nuclear magnetic resonance 
measurements (NMR) of the target material, employing a 
Q-meter~\cite{Court1993433} to observe the frequency response of an LCR circuit 
with the inductor embedded in the target material. An RF field at the proton's 
Larmor frequency induces spin flips as the proton spin system absorbs or emits 
energy. By integrating the real portion of the response as the circuit is swept 
through frequency, a proportional measure of the sample's magnetic 
susceptibility, and thus polarization, is achieved~\cite{abragam1983}. 

\begin{figure}	
  \begin{center}
    \includegraphics[width=\columnwidth]{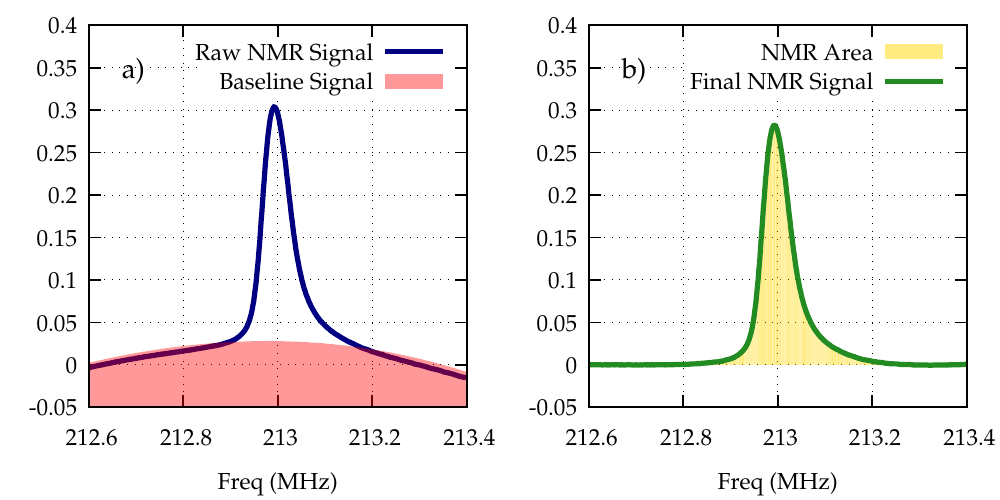}
  \end{center}
  \caption{a) Raw NMR signal and baseline in arbitrary units. b) Final NMR signal, with 
    baseline and residual signals subtracted, showing the integrated signal 
  area.}
  \label{fig:sig}
\end{figure}
NMR ``Q-curve" signals contain the frequency response of both the material's 
magnetic susceptibility, and the circuits own background response. To remove 
the background behavior of the NMR electronics, a \textit{baseline} signal is 
recorded while the proton NMR peak is shifted away from the frequency sweep 
range by lowering the magnetic field. To produce a final NMR signal, this baseline 
is subtracted, seen in a) of Figure~\ref{fig:sig}, and a polynomial fit to the 
wings of the resulting curve is performed, allowing the subtraction of any 
residual background shifts in the Q-curve, as seen in b) of Figure~\ref{fig:sig}.
 The degree of polarization is then proportional to the integrated area under this 
 background-subtracted signal.

The coefficient of proportionality used to calculate the polarization from
the integrated signal is known as the calibration constant (\textit{CC})
and is determined by NMR measurements without the application of DNP. These thermal
equilibrium (\textit{TE}) measurements provide a signal area $A_{\textrm{TE}}$ at a 
known polarization $P_{\textrm{TE}}$, calculable from the given field $B$ and temperature $T$:
\begin{linenomath} \begin{equation}
P_{\textrm{TE}} = \tanh \left(\frac{\mu B}{kT}\right).
\end{equation}\end{linenomath}
An enhanced polarization $P$ can then be calculated from a signal area $A$ during DNP:  
$P=A(P_{\textrm{TE}}/A_{\textrm{TE}})$. The calibration constant 
$P_{\textrm{TE}}/A_{\textrm{TE}}$ depends on the geometrical arrangement of 
the target material beads in the cell and the magnetic coupling of the NMR 
pickup coil to those beads, so in general a single constant may be applied to a 
target sample throughout its use in the experiment. When they were possible, 
multiple thermal equilibrium measurements for a given target material sample 
were averaged to be applied to all the target polarization data for that 
sample.

%
%

\begin{figure}	
  \begin{center}
    \includegraphics[width=0.85\columnwidth]{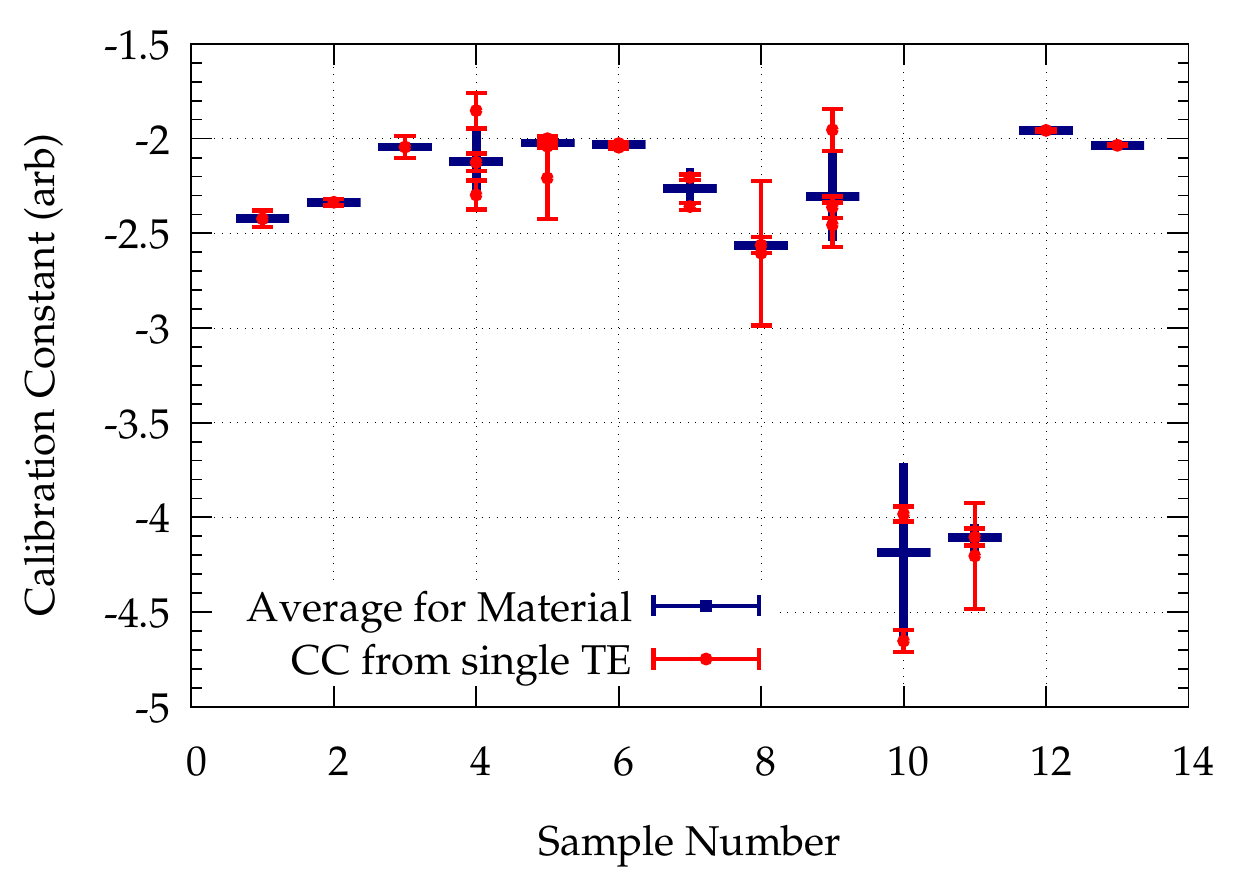}
  \end{center}
  \caption{Calibration constants for each target material sample used during 
    the experiment. The calibration constant used to calculate the final target 
    polarization is an average of one or more values from all the thermal 
    equilibrium measurements taken for that sample. Errors shown are 
  statistical only.}
  \label{fig:cc}
\end{figure}

Figure~\ref{fig:cc} shows each calibration constant taken during the experiment, 
and the final averaged constants used to calibrate the NMR signal area for each 
target material sample. Samples number 10 and 11 have drastically different 
calibration constants due to the different orientation of the NMR coil to the 
field after the magnet was rotated; they are physically the same target samples as materials 8 and 9. 

\subsection{Material Preparation and Lifetime}
%

Ammonia ($^{14}$NH$_3$) offers an attractive target material due to its high 
polarizability and radiation hardiness, as well as its favorable dilution factor --- ratio of free, 
polarizable protons to total nucleons. Ammonia freezes at 
195.5 K, and can be crushed through a metal mesh to produce beads of convenient 
size, 
allowing cooling when the material is under a liquid helium 
bath~\cite{Meyer200412}. 
%

Before dynamic polarization is possible, the material must be doped with paramagnetic 
radicals, which provide the necessary free electron spins throughout the 
material.  For SANE, the ammonia target samples were radiation doped at a small 
electron accelerator, the Medical-Industrial Radiation Facility at NIST's 
Gaithersburg campus. Free radicals were created by 19 MeV electrons at a beam 
current between 10 and 15 $\mu A$, which was incident upon the frozen ammonia 
material held in a 87 K  liquid Ar$_2$ bath, until an approximate dose of 
$100$\,Pe/cm$^2$  was achieved.  In this context, a unit of 
radiation dose of $1$\,Pe/cm$^2 = 10^{15}$ e$^-$/cm$^2$ is convenient.

While proton polarizations exceeding 95\% are possible after irradiation doping 
of ammonia, the experimental beam causes depolarization. The first depolarizing 
effect, of order 5\%, is due to the decrease in DNP efficiency due to 
excess heat from the beam~\cite{Liu19981}. A longer term depolarization effect 
comes from the build up of excess radicals under the increasing dose of 
ionizing radiation. These excess radicals mean more free electrons which 
provide more paths for proton relaxation and depolarization.

\begin{figure}		
  \begin{center}
    \includegraphics[width=0.92\columnwidth]{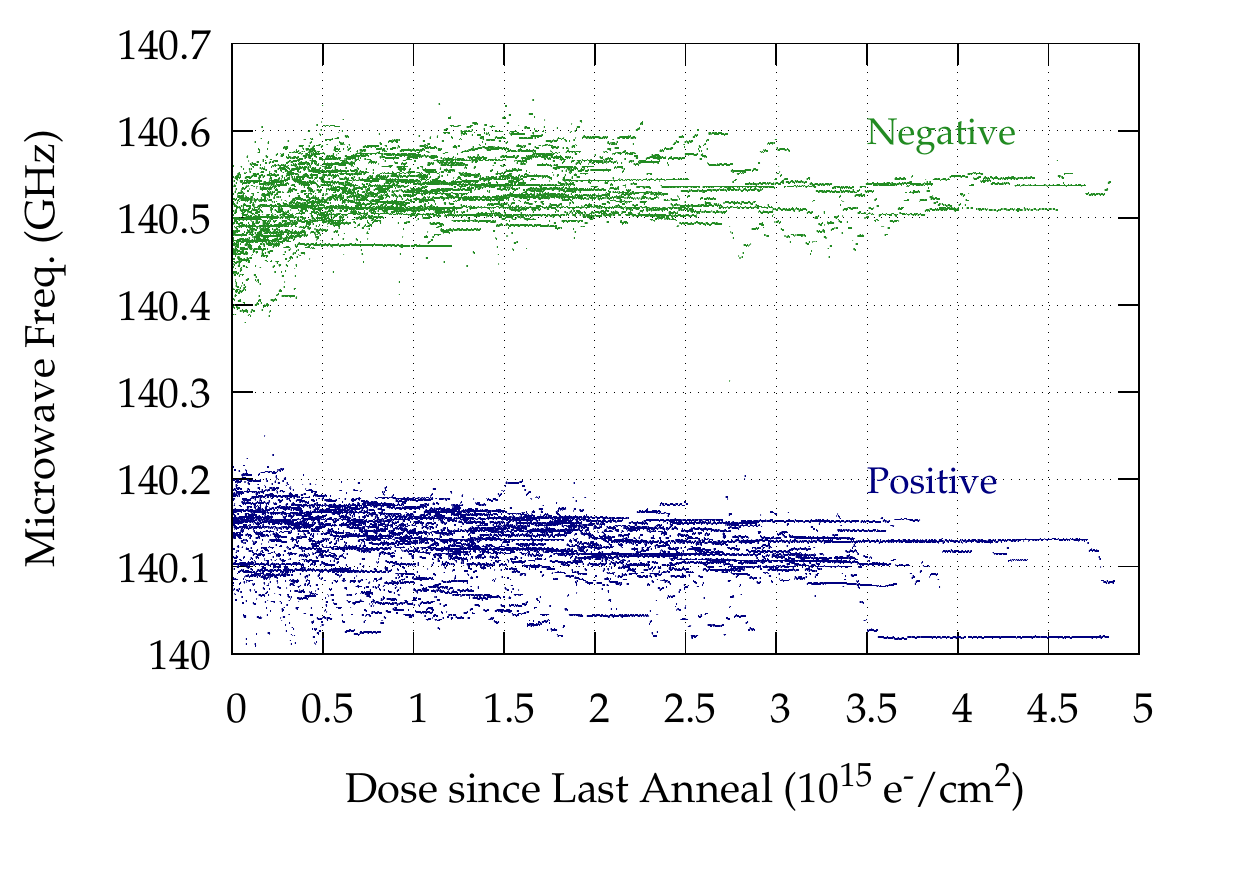}
  \end{center}
  \caption{The change in microwave frequency used to polarize during SANE as radiation dose 
    from the beam is accumulated. Positive polarization points (below 140.3\,GHz) 
   show a roughly linear decrease, while the negative polarization points  (above 140.3\,GHz) exhibit a curving increase.}
  \label{fig:uwave}
\end{figure}

By heating the target material to between 70 and 100\,K, certain free radicals can 
be recombined. This \textit{anneal} process will often allow the polarization to 
achieve its previous maximal values. With subsequent anneals, however, the 
build-up of other radicals with higher recombination 
temperatures will result in an increased decay rate of the polarization, until 
the material must be replaced~\cite{McKee200460}.

While the maximum achievable polarization
falls as continued radiation dose is accumulated, the optimal microwave 
frequency needed to reach the highest polarization will also shift as the free 
electrons come under the dipole--dipole influence of more free electron 
neighbors, broadening the electron spin resonance peak.  Figure~\ref{fig:uwave} 
shows the shift in microwave frequency chosen by the target operator during the 
experiment, as a function of the dose accumulated on the target since the last anneal.

\begin{figure}	
  \begin{center}
    \includegraphics[width=0.9\columnwidth]{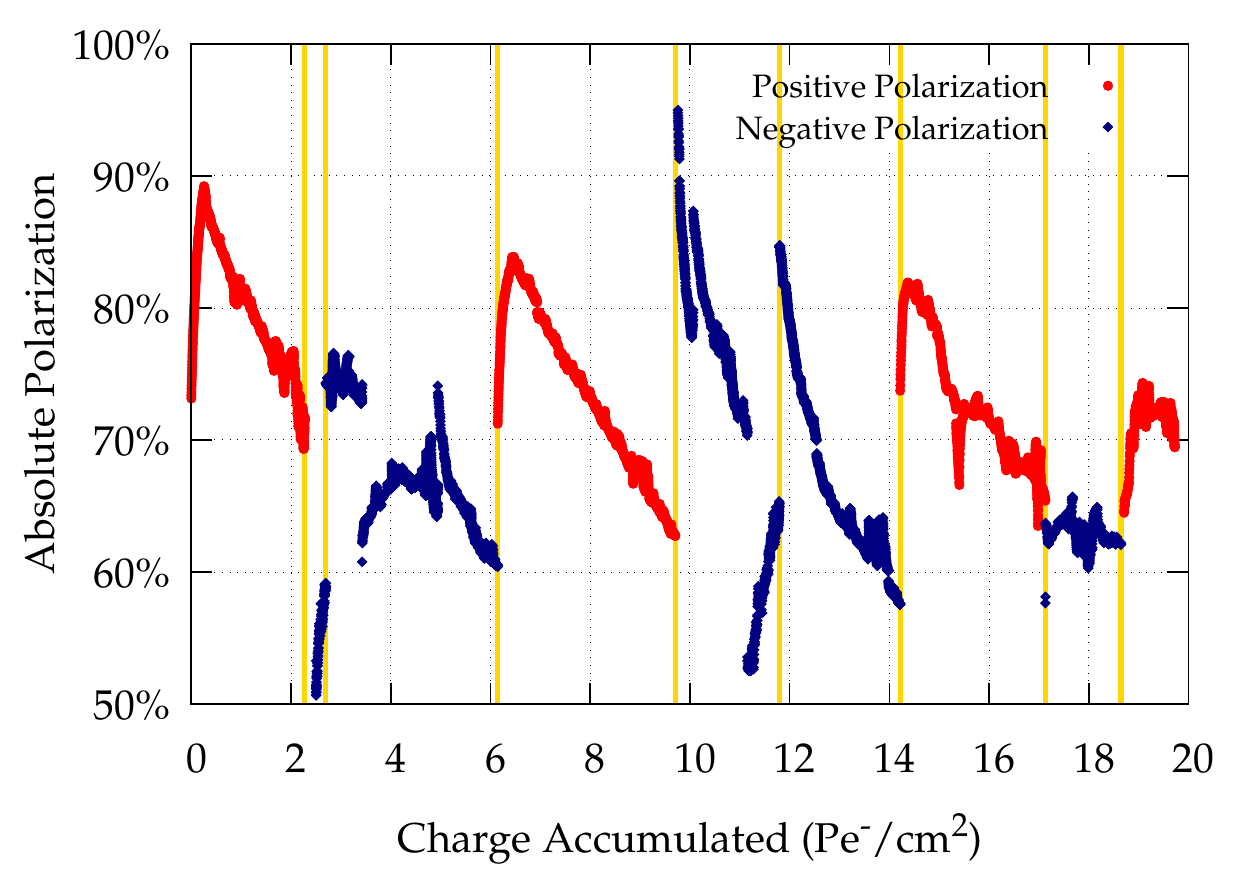}
  \end{center}
  \caption{Polarization of a typical target material sample versus charge 
    accumulated during data taking, with vertical yellow lines showing when 
  anneals were performed.}
  \label{fig:life}
\end{figure}

Figure~\ref{fig:life} shows the lifetime of a typical target material used 
during SANE, and illustrates several artifacts common during beam taking 
conditions. Vertical yellow lines depict anneals. The build-up of radicals in 
beam can be seen at 0 and 6\,Pe/cm$^2$ as polarization actually increases with 
dose accumulated. Small spikes in polarization seen throughout are the result 
of beam trips, when the polarization improves as the temperature drops with the 
loss of heat from the beam. Other hiccups in operation apparent in the plot are 
a poorly performed anneal, just after 2\,Pe/cm$^2$, resulted in starting 
polarization below 60\%, and the loss of liquid helium in the target cell at 
approximately 3 and 11\,Pe/cm$^2$.

\subsection{Offline Corrections}

Several corrections were necessary to the online NMR signal analysis that was 
performed as the experiment ran. Because the scale of the thermal equilibrium 
signals is two orders of magnitude smaller than that of the enhanced 
polarization signal, different amplification gains are used for the two measurements.  
Differences between the nominal and actual gains of the amplifiers result in a 
correction of approximately 1\%.

During the running of the experiment, the superconducting magnet experienced a damaging quench which necessitated repairs. While 5\,T operation of the magnet was restored, a slight current leak while in persistent mode was seen due to minute electrical resistance~\cite{Maxwell:2017ubl}. While the change in magnet current was 
only about 0.05\% per day, this resulted in a significant shift in the NMR signal 
peak. The wings of each signal---after baseline subtraction--- are used to 
perform a polynomial fit to remove residual Q-curve movement, so the shifting 
peak created poor fits as it approached the edge of the sweep range.  
This effect was corrected by varying the size of the wings used in the 
polynomial fit for each signal, ensuring that only the background portion of 
the signal was included in the fit.


\subsection{Target Polarization Performance}

\begin{figure}	\begin{center}
    \includegraphics[width=0.9\columnwidth]{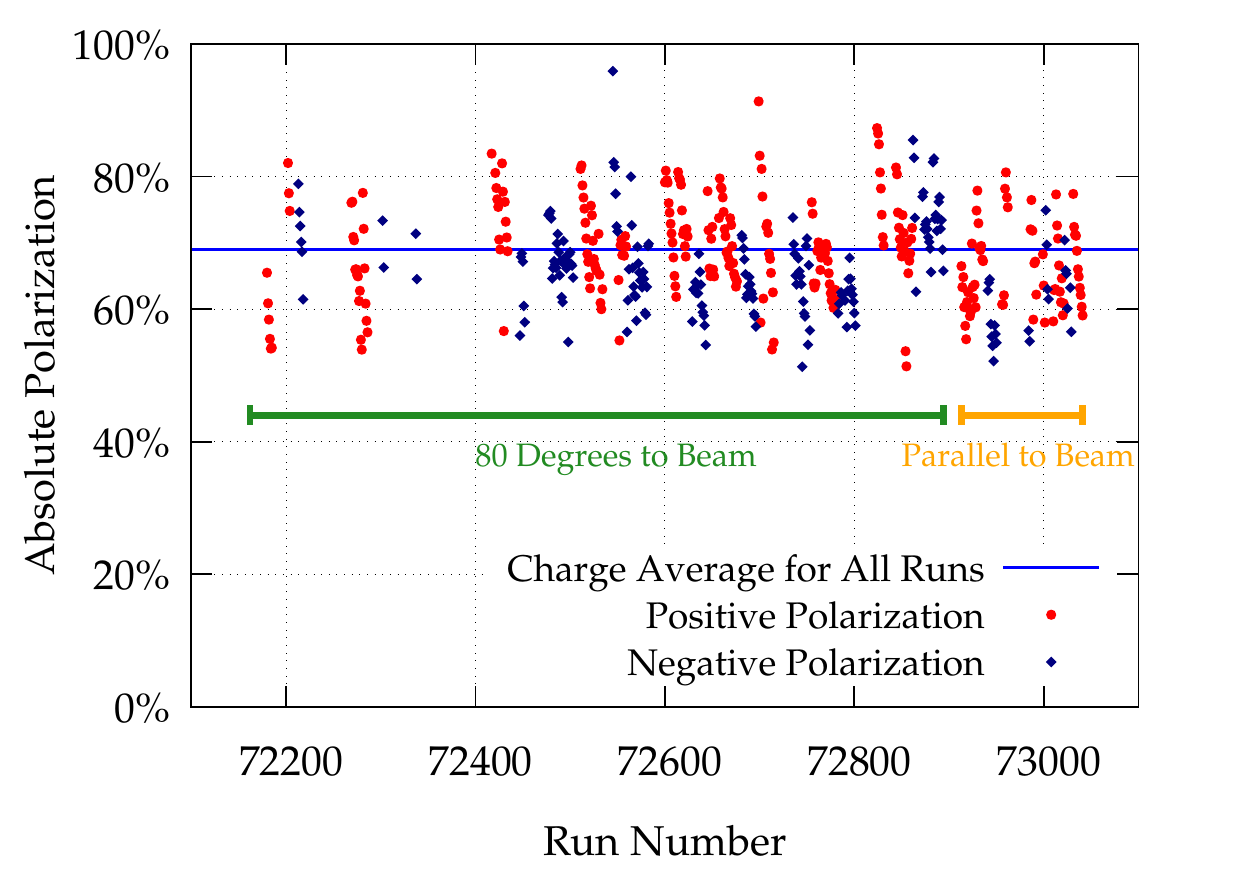}
  \end{center}
  \caption{Charge averaged target polarization achieved for each SANE data-taking run.}
  \label{fig:tar_pol}
\end{figure}

During SANE, a total of 122.2\,Pe/cm$^2$ of radiation dose was accumulated on the 11 different ammonia material samples.  Anneals were 
performed 26 times, and 23 thermal equilibrium calibration measurements were taken.  Figure~\ref{fig:tar_pol} shows the polarization for each experimental run, with 
indications for the orientation of the target during that period. Despite 
considerable unforeseeable difficulties in the operation of the target during 
SANE, the total charge-averaged proton polarization achieved was 68\%.

\section{Detector Systems} 

\begin{figure}
	\begin{center}
		\includegraphics[width=\columnwidth]{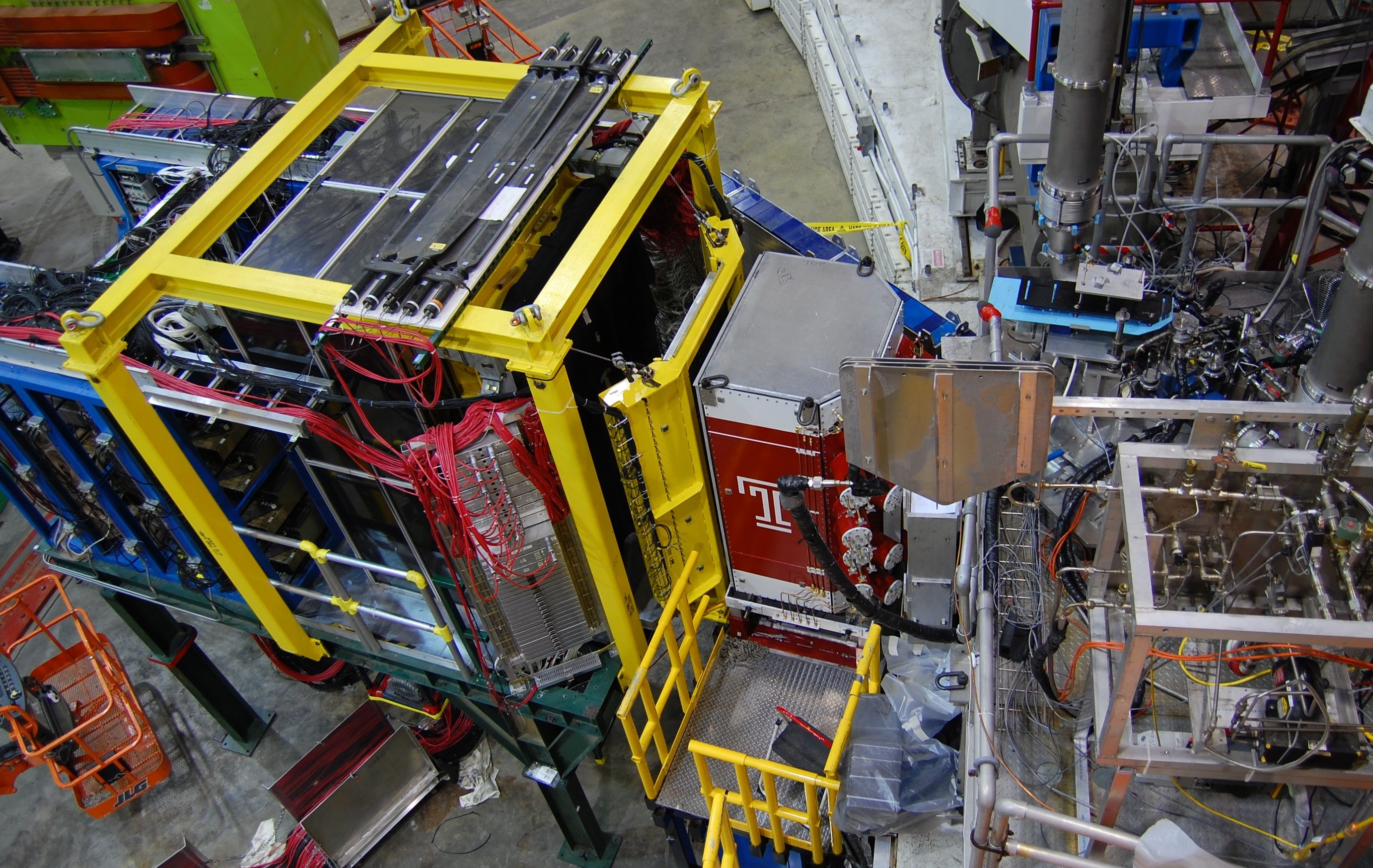}
	\end{center}
	\caption{Photograph of BETA from above, showing the support structure for the calorimeter at left, lucite hodoscope in yellow at center, Cherenkov tank in red, and target platform at right.}
	\label{fig:betaphoto}
\end{figure}

\begin{figure}
  \begin{center}
    \includegraphics[width=\columnwidth]{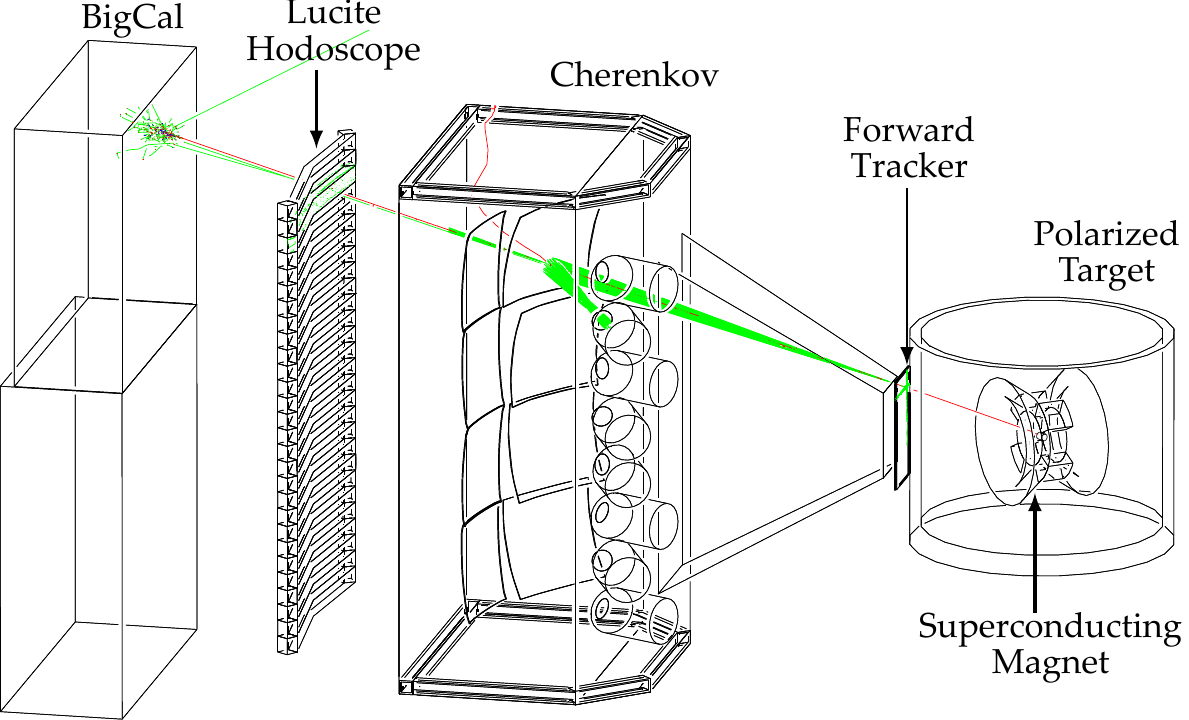}\\
    \vspace{0.1in}
    \includegraphics[width=\columnwidth]{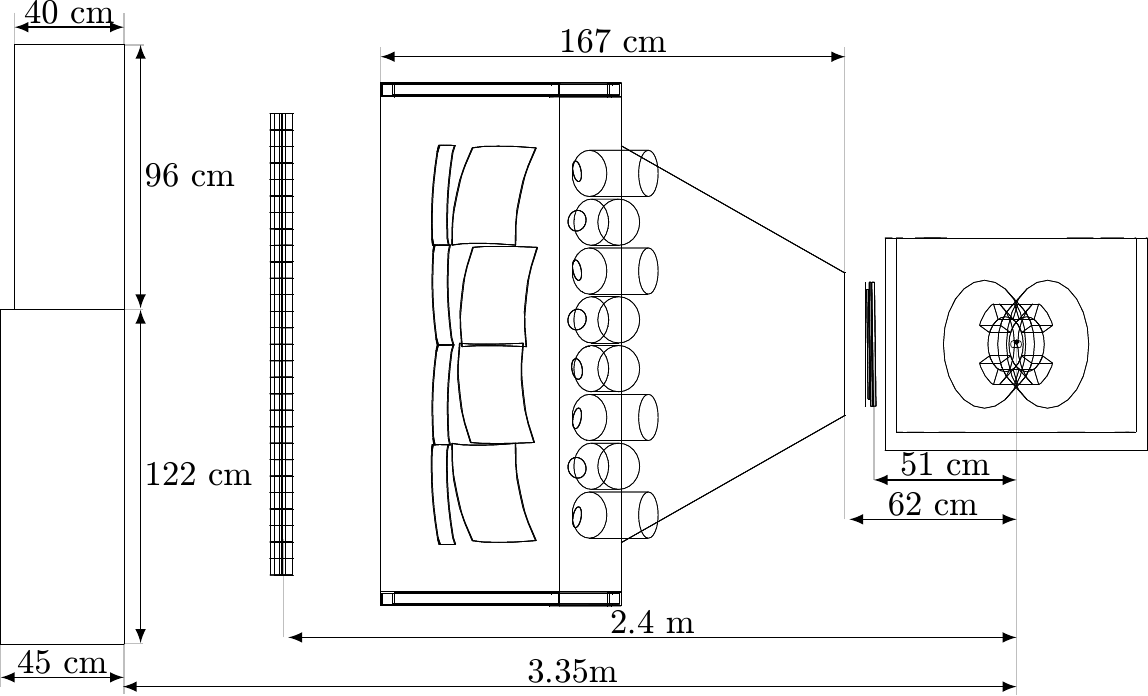}
  \end{center}
  \caption{Two renderings of BETA from the Geant4 simulation, showing at top a simulated electron 
    event originating in the target, creating Cherenkov showers in the gas 
    Cherenkov and lucite hodoscope, and depositing its energy in the upper 
  section of the calorimeter. The lower diagram shows the dimensions of each components, and their distances from the target.}
  \label{fig:beta}
\end{figure}

The centerpiece of SANE's inclusive measurement of deep inelastic electron 
scattering was the Big Electron Telescope Array (\textit{BETA})\footnote{The original BETA design was conceived by Glen Warren \cite{warren}.}, a large 
acceptance, non-magnetic detector package situated just outside the target 
vacuum chamber (see Figure~\ref{fig:betaphoto}). Electrons scattered in the target passed though a small 
tracking hodoscope for position information, a threshold Cherenkov detector for 
electron discrimination, and a second, large hodoscope, before finally 
producing a shower in the calorimeter. BETA occupied a large, 0.2\,sr solid 
angle at 40\degrees~to the beam direction, and provided pion rejection of 
1000:1, energy resolution of better than $10\%/\sqrt{E}$, and angular 
resolution of approximately 1\,mr. Figure~\ref{fig:beta} shows renderings
 of a Geant4 simulation of BETA with an example 
electron track.

%

%

\subsection{BigCal} 

\begin{figure}
  \begin{center}
    \includegraphics[width=\columnwidth]{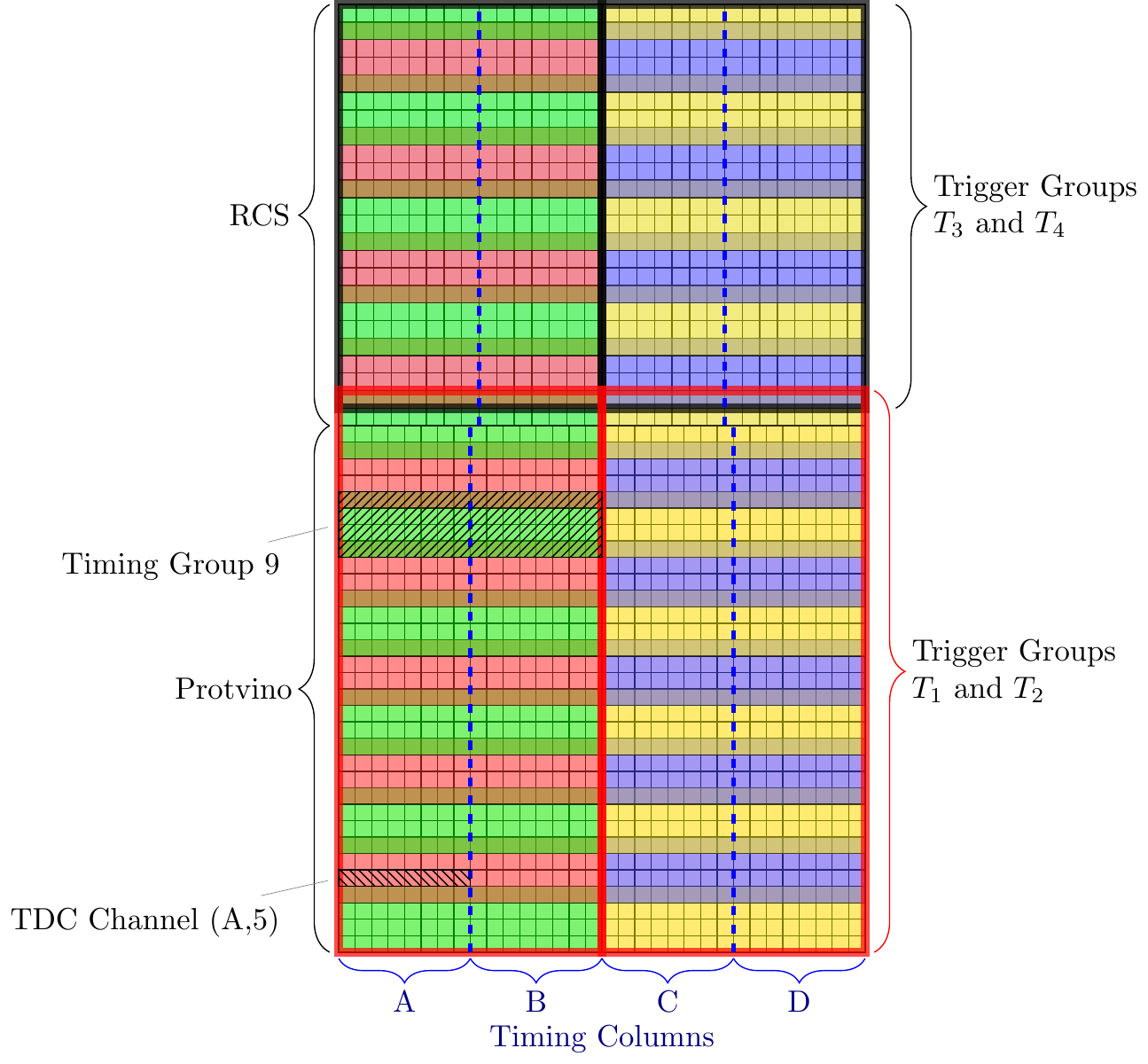}
  \end{center}
  \caption{Layout of BigCal's 1,744 lead-glass blocks, showing upper RCS and 
    lower Protvino sections, as well as trigger and timing groups. An example 8 
    block TDC channel and 64 block timing group are show in hatched 
  areas~\cite{Armstrong:2015ukn}.}
  \label{fig:bigcal}
\end{figure}

BETA’s big electromagnetic calorimeter, \textit{BigCal}, consisted of 1,744 
TF1-0 lead-glass blocks; 1,024 of these were 3.8 $\times$ 3.8 $\times$ 45.0 
cm$^{3}$ blocks contributed by the Institute for High Energy Physics in 
Protvino, Russia. The remaining 720, from Yerevan Physics Institute, were 4.0 
$\times$ 4.0 $\times$ 40.0 cm$^{3}$ and were previously used on the RCS 
experiment~\cite{PhysRevLett.94.242001}. The calorimeter was assembled and 
first utilized by the GEp-III collaboration 
\cite{Puckett}. The Protvino blocks were 
stacked 32 $\times$ 32 to form the bottom section of BigCal, and the RCS blocks 
were stacked 30 $\times$ 24 on top of these, as seen in Figure~\ref{fig:bigcal}.  The 
assembled calorimeter had an area of roughly 122 $\times$ 218 cm$^{2}$, which, 
placed 335\,cm from the target cell, made a large solid angle of approximately 
0.2 sr at a central scattering angle of 40\degrees.

BigCal was the primary source for event triggers for BETA, and a summation 
scheme was used to simplify triggers and reduce background events, summarized in 
Figure~\ref{fig:bigcal}. While each lead-glass block had its own FEU-84 
photomultiplier tube and ADC readout, the smallest TDC readouts consisted of 
groups of 8 blocks in one row. These TDC groups then formed 4 timing columns, 
which were summed and discriminated for another TDC readout. The 8 block TDC 
signals were also summed into larger timing groups of 64 blocks, 4 rows by 8 
columns (designated by color in Figure~\ref{fig:bigcal}), which were 
overlapped to avoid split events. Finally, timing groups were summed into four 
trigger groups to form the main DAQ triggers~\cite{Puckett}.

\subsection{Gas Cherenkov}

The Cherenkov counter held dry N$_{2}$ radiator gas at near atmospheric 
pressure, and employed eight 40 $\times$ 40 cm$^{2}$ mirrors to focus Cherenkov 
photons onto 3 inch diameter Photonis XP4318B photomultiplier tubes. Nitrogen's index of 
refraction of 1.000279 gave a momentum threshold for Cherenkov emission by 
pions of 5.9\,GeV/c, allowing effective rejection of pions, given a maximum beam 
energy of 5.9\,GeV. The 8 mirrors, 4 spherical and 4 toroidal, were positioned 
to cover the full face of BigCal, effectively dividing BigCal into 8 geometric 
sectors each corresponding to one mirror. Due to the proximity of the Cherenkov 
tank to the target magnetic field, $\mu$-metal shields enclosed each 
photomultiplier tube, and additional iron plating was situated between the tank 
and magnet. The design and performance of the SANE Cherenkov is 
discussed in detail in reference \cite{Armstrong}.

\subsection{Hodoscopes}

Two tracking hodoscopes provided additional position information and background 
rejection. Mounted between BigCal and the Cherenkov tank, the lucite 
hodoscope consisted of 28 lucite bars of 3.5 $\times$ 6.0 $\times$ 80.0 cm, 
curved with a radius equal to the distance from the target cell, giving a 
normal incidence for participles originating in the target. With an index of 
refraction of 1.49, Cherenkov radiation was produced from the passage of 
charged particles above $\beta_{\text{threshold}} = 0.67$. The effective threshold increases to 0.93
 when Cherenkov photons are detected simultaneously at both ends of the lucite bar, because these photons propagate through total internal reflection. The Cherenkov angle must be above critical angle for lucite  (42\degrees) in this case. Photonis XP2268 
photomultiplier tubes coupled to the end of each bar collected the 
Cherenkov light, allowing the determination of the position of the hit along 
the bar using timing information from both tubes.

A smaller, front tracking hodoscope consisted of three planes of  3 $\times$ 3 
mm Bicron BC-408 plastic scintillator bars positioned just outside the target 
scattering chamber, 48\,cm from the target cell. This hodoscope provided 
tracking information on particles as they were still under the influence of the 
target's magnetic field. By combining tracking information close to the target 
with final positions in BigCal, any discernible curve in the particles 
trajectory would allow differentiation of positively and negatively charged 
particles, allowing positron rejection.

\subsection{Hall C HMS}

The standard detector system in  Hall C, the High Momentum Spectrometer (HMS), 
was utilized in a supporting role throughout the experiment. The HMS is made up 
of three superconducting quadrupole magnets and one superconducting dipole, 
which focus and bend charged particles into a detector package with two gas 
drift chambers, four hodoscopes, a gas Cherenkov tank and a lead-glass 
calorimeter. During SANE, the HMS was positioned at 15.4\degrees, 16.0\degrees~ 
and 20.2\degrees, accepting proton and electron scattering events from the 
target. In addition to the calibration and support of BETA, events from the HMS 
were used to produce independent analyses on the proton electric to magnetic 
form factor ratio~\cite{anusha} and spin asymmetries and structure 
functions~\cite{kang}.

\subsection{Data Acquisition}

Data collection was coordinated by a trigger supervisor~\cite{jastrzembski}, 
which received triggers from BigCal, Cherenkov and HMS TDCs. If not busy, the 
trigger supervisor accepted triggers from readout controllers, sending gate 
signals to ADCs and start signals to TDCs. Readout controllers then read out 
signals, to be assembled by an event builder and saved to disk. To monitor 
events missed due to the data acquisition being in a busy state, the dead-time 
was monitored with scalers on the discriminator output which wrote to the data 
stream every 2 seconds. 

SANE utilized 8 trigger types, representing triggers and coincidences from the 
detectors, of which 2 were used in the final analysis. The \texttt{BETA2} 
triggers were the result of coincident hits in the Cherenkov and BigCal, 
representing a candidate electron event.  \texttt{PI0} triggers required two 
BigCal hits in different quadrants of the detector, representing  two, vertically-separated  photon events from neutral pions.

\section{BETA Commissioning and Calibration}

SANE's initial commissioning and calibration schedule was interrupted by an unanticipated
target magnet failure and subsequent repairs. The delays meant the cancellation of plans  to calibrate BigCal with elastic $e$-$p$ scattering using coincidences with protons detected in the HMS. In this scheme, the target magnet strength and orientation would
have been varied to scan the elastic events across the full face of the calorimeter 
while running at reduced beam energy. 
In order to optimize data collection for the 
proposed beam energy and target configurations while accommodating the 
accelerator run plan, the commissioning of the BETA detectors began with transverse  target magnet orientation rather than parallel.
In total, the target magnet failure and unrelated accelerator operation issues contributed to 
roughly 45\% fewer data being collected than was originally proposed.

Instead, BETA's BigCal calorimeter was calibrated in real-time using neutral 
pion events from the target, allowing drifts in gain to be observed throughout 
the experiment. The Cherenkov photomultiplier tube ADC channels were 
calibrated before the experiment to roughly 100 channels per photo-electron, as 
discussed in detail in reference \cite{Armstrong}. The Lucite hodoscope was used 
only for TDC data to record the position of hits, calculable from propagation of 
the electron's Cherenkov light to photomultiplier tubes at each end of the 
bar.

\subsection{Cluster Identification}

To reconstruct the final energy and position of particle hits in the 
calorimeter, a simple algorithm was used to group signals originating from one 
shower in neighboring calorimeter blocks  into clusters for each event. The 
block with the largest signal was selected as the cluster seed, and struck blocks 
within a 5$\times$5 grid of this centroid were included in the cluster, unless  
detached from the group. The next cluster was formed by finding the next 
highest signal block, excluding those already included in a cluster, and this 
process was repeated until all blocks above a chosen threshold  were 
used.

Once clusters were identified, they were characterized for use in the analysis. 
We assigned each cluster a pre-calibration energy $E_c=\sum_{i}c_iA_i$ for 
block number $i$, ADC values $A_i$ and block calibration constants $c_i$, where 
final $c_i$ are the end goal of the calibration. In the first pass of analysis, 
each ADC channel was assumed to be 1 MeV, based on adjustments before the 
experiment using cosmic ray events. The moment of the cluster is then an energy 
weighted average of position
\begin{linenomath} \begin{equation}
\langle x\rangle = \sum_i \frac{c_iA_i}{E_c}(x_i-x_{\textrm{seed}}),
\end{equation}\end{linenomath}
and similarly for $\langle y\rangle$, so that the cluster position on the face 
of BigCal was taken to be $(x_{\textrm{seed}}+\langle 
x\rangle,y_{\textrm{seed}}+\langle y\rangle)$. The second moment gave the 
position standard deviation.

\subsection{$\pi^0$ Calibration}

The large number of $\pi^0$ background events incident on the calorimeter from 
the target allowed reliable calibration of a majority of the calorimeter, as 
well as effective, real-time gain monitoring throughout the experiment. Neutral 
pions produced in the target decay to two photons at a 98.8\% branching 
probability with a mean lifetime of $8\times10^{-17}$ seconds, so that most 
pions have decayed to photons before exiting the target. By measuring the 
separation angle of the photons $\alpha$, we can determine the relative 
energies of the incident photons $E_{1,2}$ from the pion mass $m^2_{\pi^0}= 2 
E_1E_2(1-\cos\alpha)$.

Unfortunately, the \texttt{PI0} trigger was unable to populate all calorimeter 
blocks with events because the trigger required two of the four trigger groups 
 to fire in coincidence ($T_{1-4}$ shown in Figure~\ref{fig:bigcal}). The reach 
of the events was limited by the energy thresholds for each trigger groups' 
discriminator, which was set to roughly 400 MeV. For example, to populate the 
upper-left most block with a photon shower requires relatively low energy 
$\pi^0$ decays, so that the angle between the two photons is large enough to 
trigger $T_3$ and $T_4$. If the $\pi^0$ is too energetic, the angle is not big 
enough to reach both trigger groups. In hindsight, the solution would have been to use 
smaller trigger groups to form the \texttt{PI0} trigger. 

To supplement the $\pi^0$  calibration
and improve the energy calibration of blocks at the edges of the calorimeter, 
a calibration  was done
by looking at the energy spectra measured in each block.
A GEANT simulation of the experiment was run with events weighted by the 
inelastic cross section~\cite{pbec_F1F21}. The energy spectra
for each block is dominated by inelastic electrons in the
high energy tail. 
The energy gain coefficients for a block were set so that
the measured energy spectra for each block matched
the GEANT simulated energy spectra in the high energy tail region
for $W < 2.0$\,GeV. These energy gain coefficients were used as the
starting values for 
determining the final gain coefficients in the  $\pi^0$  calibration method.

\begin{figure}[htb]
  \begin{center}
      \includegraphics[width=0.8\columnwidth, trim=14mm 3mm 2mm 
      8mm,clip]{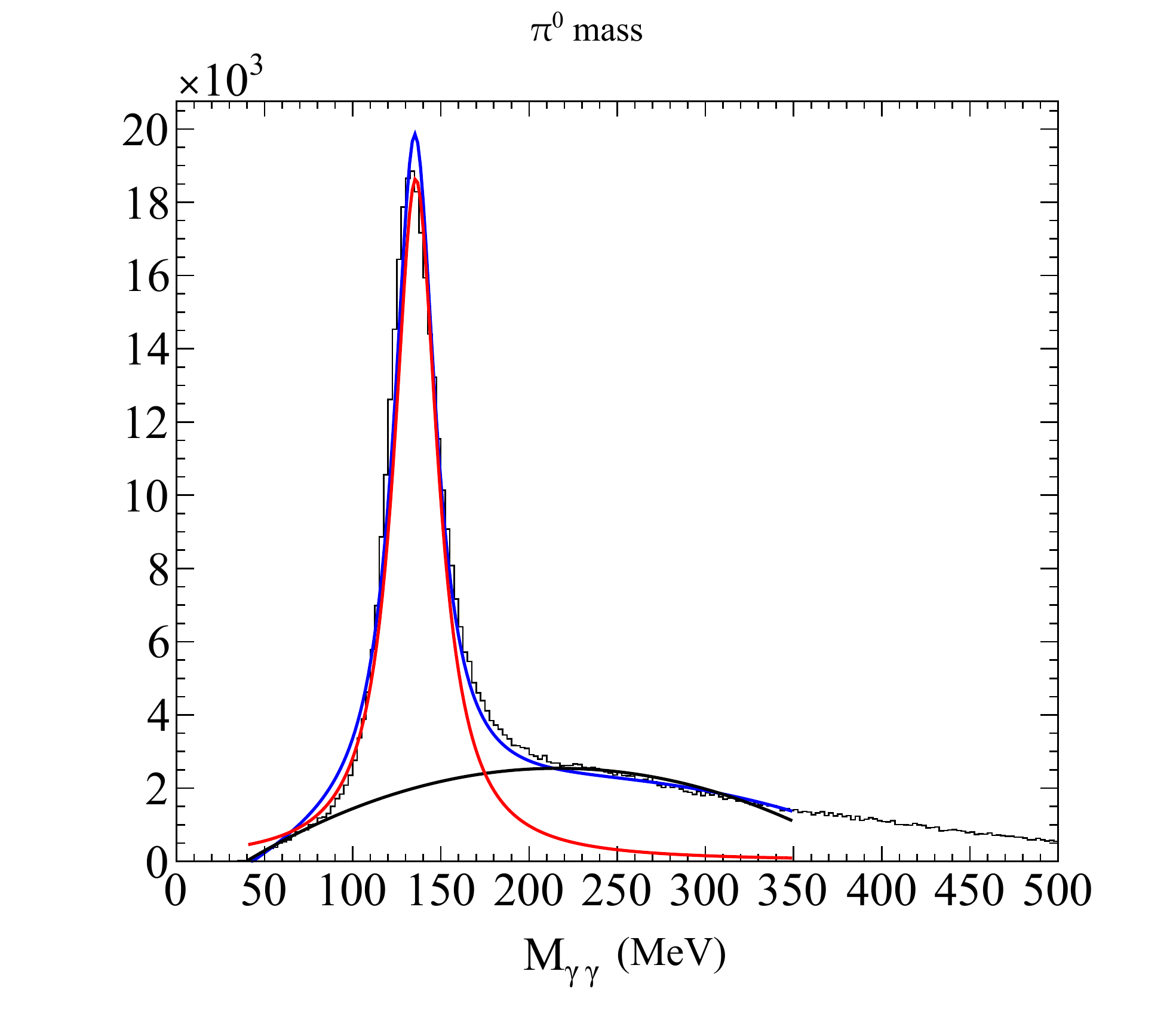}
  \end{center}
  \caption{Plot of neutral pion mass reconstruction after block calibration.  
    The energy resolution of this peak is directly proportional to the energy 
  resolution of the clusters in the calorimeter.}
  \label{fig:mpi}
\end{figure}

Events from the \texttt{PI0} trigger were chosen and cuts were placed to 
include only clusters which were 20\,cm to 80\,cm apart, excluding pairs 
produced outside the target, and to exclude events that gave triggers in the 
Cherenkov, such as electrons. To calibrate a given block, a 
histogram of the invariant mass results was formed for all the clusters which passed the cut 
and included that block. Normalizing this invariant mass result to the known 
pion mass $\pi^0=134.9$\,MeV, a new calibration constant was obtained for the 
block. Once new constants were produced for all blocks, this process was 
repeated and iterated many times until all block results converged on the pion 
mass, as seen in Figure~\ref{fig:mpi}.

\begin{figure}[htb]
  \begin{center}
    \includegraphics[width=0.8\columnwidth]{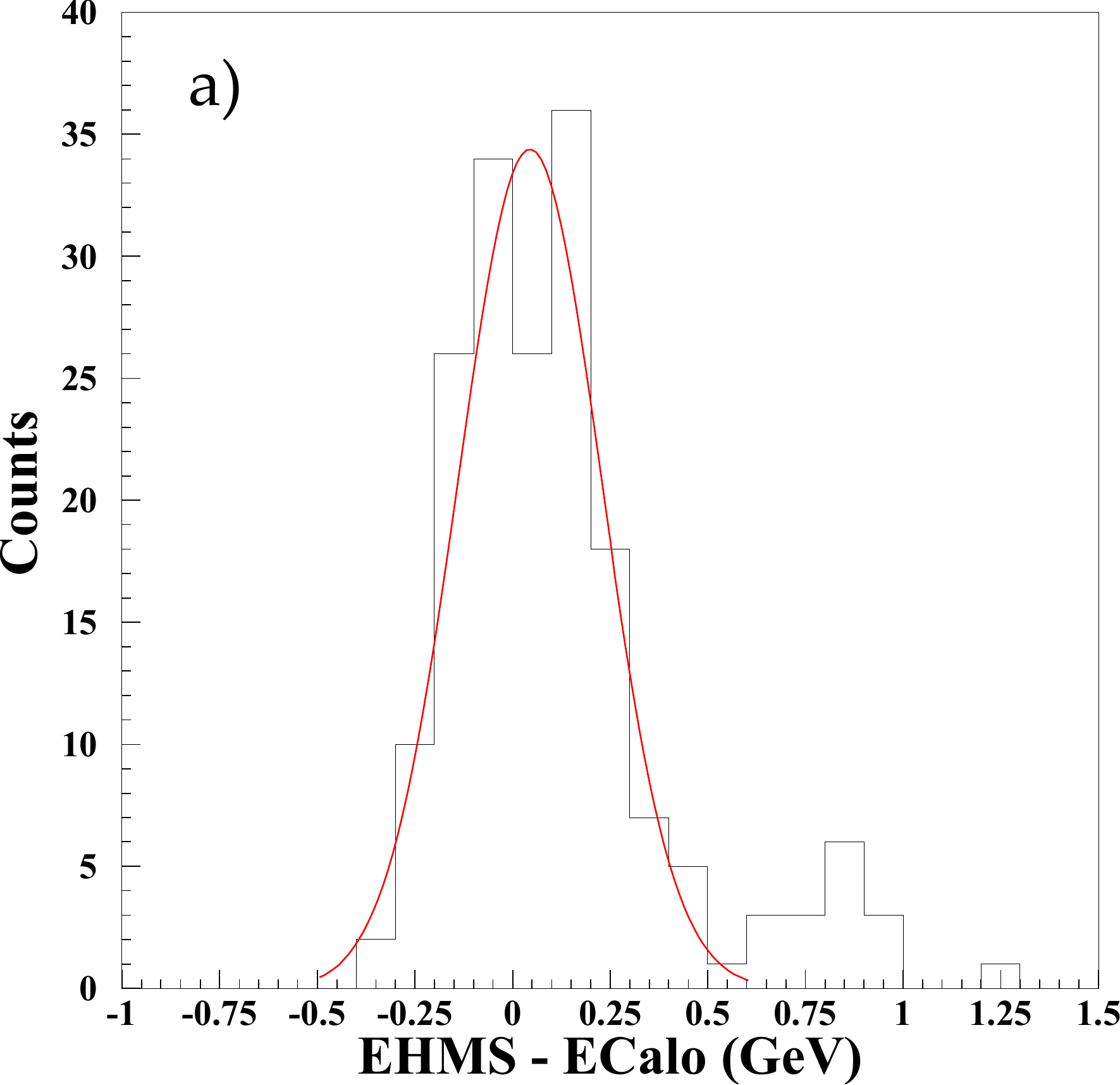}\\
    \vspace{0.1in}
    \includegraphics[width=0.8\columnwidth]{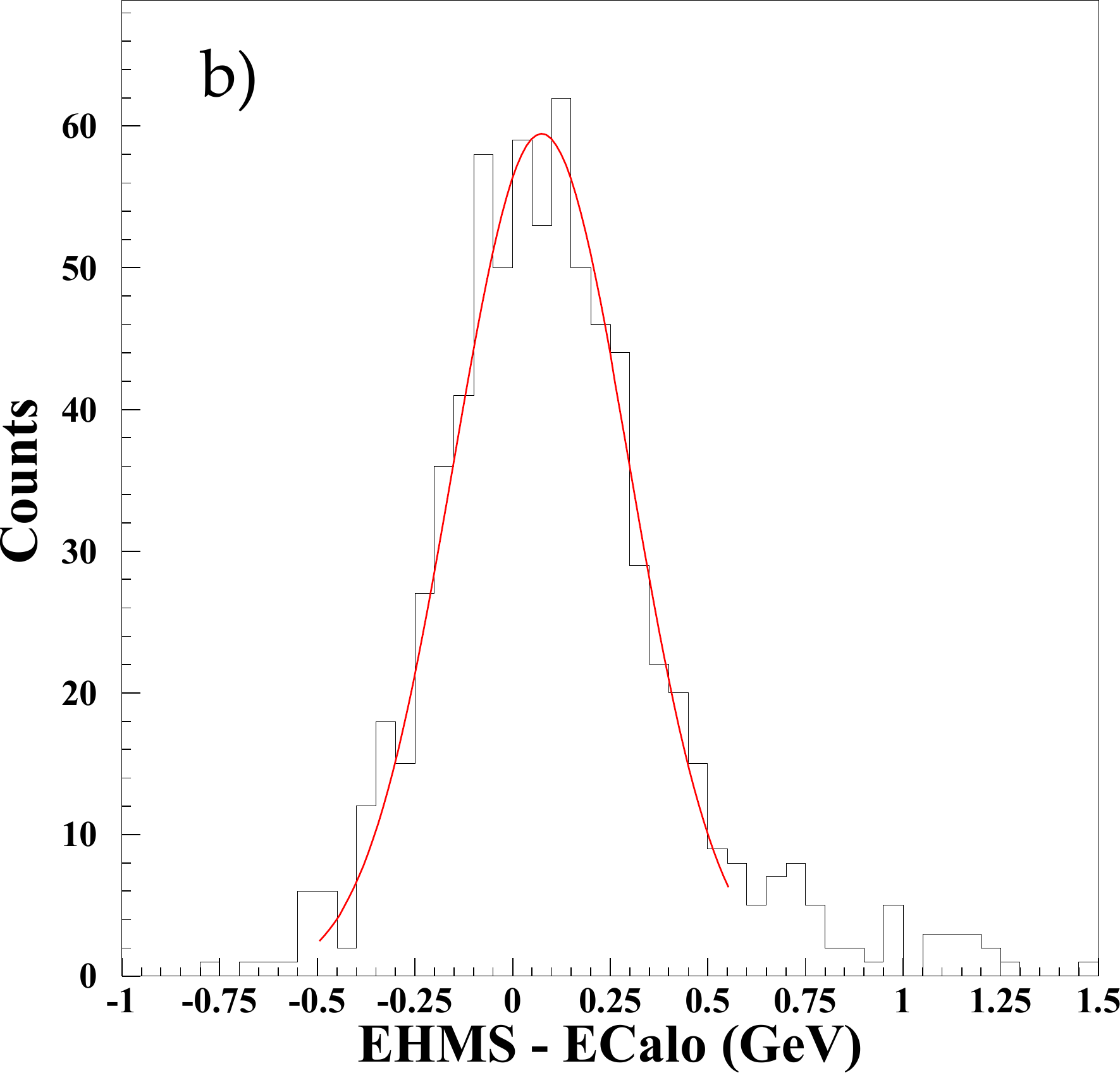}
    
    \caption{\label{fig:betaElastic}The difference of electron energies reconstructed from elastic protons detected in the HMS and the measured energies in BETA for 4.7\,GeV (a) and 5.9\,GeV (b) beam energies. }
  \end{center}
\end{figure}

Simultaneous with the collection of BETA's main inclusive $e$ data, $e$-$p$ elastic
 coincidence data was taken employing the HMS to 
 gather the proton's momentum and angle.
Using the known beam energy and the measured proton momentum in the HMS, the scattered electron energy  can be calculated (EHMS), giving the only explicit measure of the calorimeter
energy resolution for electrons.
 The acceptance-averaged value of the
 electron momentum was 2.0 and 2.6\,GeV for beam energies of 4.7 and 5.9\,GeV.
The difference between EHMS and the energy
measured in the calorimeter (ECalo) is plotted in Figure~\ref{fig:betaElastic} for the beam energies of 4.7 (a) and 5.9 GeV (b); Gaussian fits show energy resolutions of 9.1 $\pm$ 0.5\% and 9.08 $\pm$ 0.03\% in each case.


\subsection{Neural Networks and Track Reconstruction}

Three neural networks were constructed to aid the track reconstruction for BETA: (a) a BigCal position 
correction network, which determined the $x$--$y$ coordinate where the a photon track crossed the calorimeter 
face; (b) a second network for the $x$--$y$ coordinate correction for charged particles, necessitated by the 
difference between the shower profiles of electrons and positrons, and photons; and (c) a network to determine the 
scattered momentum vector at the target, correcting for the deflection of 
charged tracks as they propagated through the target magnetic field. Each neural 
network was trained for each particle type (electron, positron, and photon) and 
target field/beam energy configuration.
 A Geant4 simulation with a detailed description of the geometry and an extended 
target field map was used to generate the events for training each neural 
network. Roughly 1 million events were simulated with uniformly distributed angle and energy, and originating uniformly from the target volume.

\subsubsection{Photon Position Corrections} \label{sec:posnn}

Particles incident on the calorimeter farther away from the center of its face 
arrived at more oblique angles to the surface, so that the depth of the shower 
had an increasing effect on the resolved cluster moment.
Photons hitting the calorimeter at the top or bottom enter the face of the 
calorimeter at angles far from normal incidence. Therefore the electromagnetic 
shower's longitudinal development will have the same directional bias.  The x 
and y moments for these types will result in a shift that depends on the 
incident angle (which for photons is easily mapped to its position).
In order to correct for this, a neural network (a) was trained to provide the 
reconstructed x-y coordinates of where the photon crossed the face of the 
calorimeter.
The neural network provided the correction values $\delta_{x}= 
x_{\textrm{face}}-x_{\textrm{cluster}}$ and $\delta_{y}= 
y_{\textrm{face}}-y_{\textrm{cluster}}$, the difference between the position on 
the face of BigCal where the particle entered and centroid of the cluster 
created in BigCal.

\begin{figure}[htb]
	\begin{center}
		{%
			\includegraphics[width=0.8\columnwidth, trim=14mm 3mm 10mm 13mm,clip 
			]{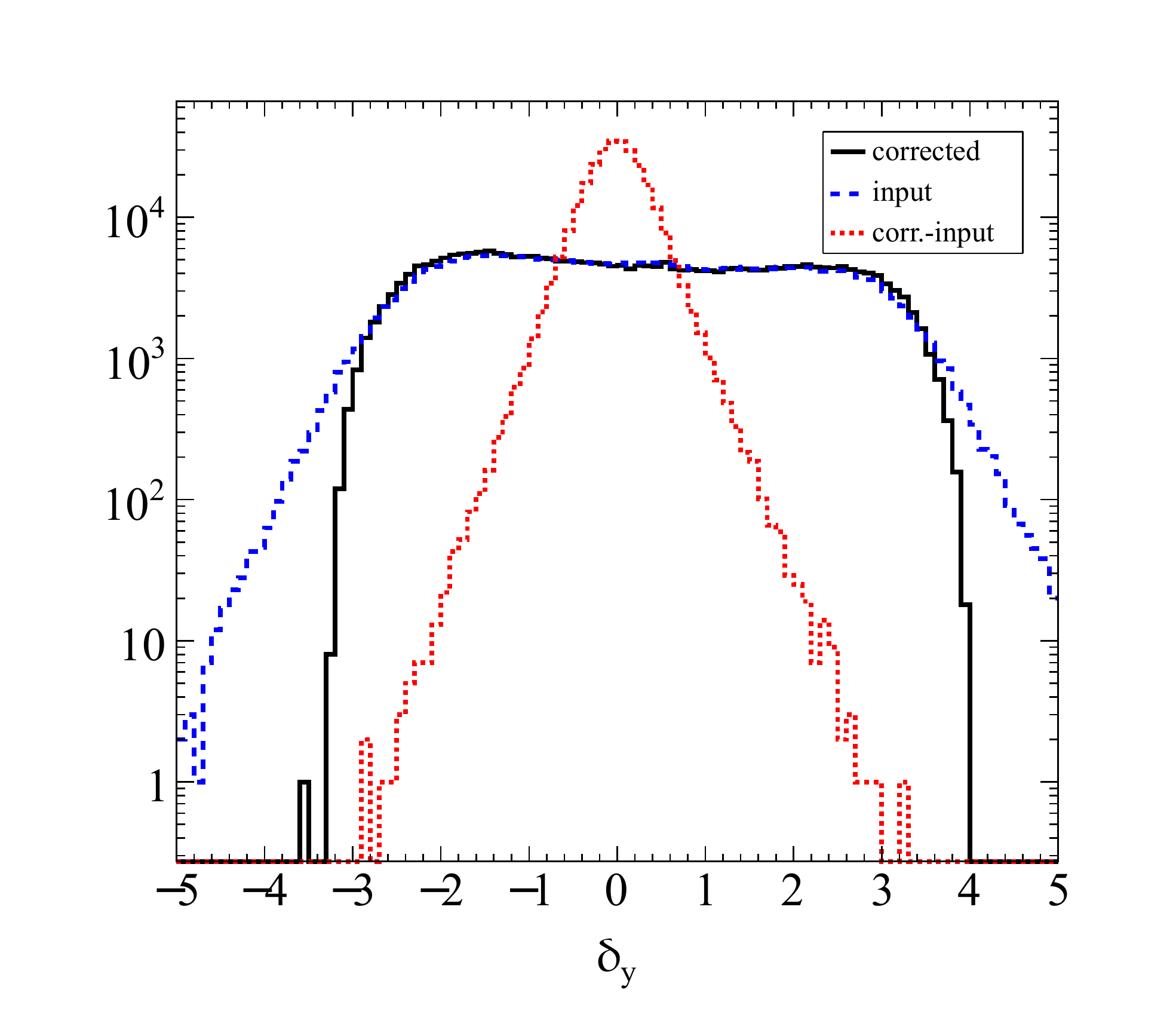}
		}
	\end{center}
	\caption{\label{fig:nn}The performance of the network correction on the cluster $y$ 
		position (in cm). The blue (long dash) histogram shows the simulation input data used to 
		train the network. The black (solid) histogram shows the network result. The red (small dash) 
		histogram shows the difference between the two.}
\end{figure}

This photon position correction neural network (a) followed the Broyden-Fletcher–Goldfarb–Shanno (BFGS) training method 
\cite{Byrd}, using a sigmoid activation for all nodes. Quantities characterizing the 
cluster, such as its mean position, standard deviation, skewness and kurtosis, 
were used as input neurons. The strongest neuron weights for the $\delta_y$ 
correction were connected to the $y$ position input neuron, so that with increasing
distance from the calorimeter center, the correction 
for the oblique angle of incidence increased, as well.  Figure~\ref{fig:nn} 
shows the performance of the neural network for the $y$ position correction.  

\subsubsection{Electron Reconstruction} 

\begin{figure}[htb]
  \begin{center}
        \includegraphics[width=0.8\columnwidth, trim=0mm 0mm 0mm 0mm,clip 
        ]{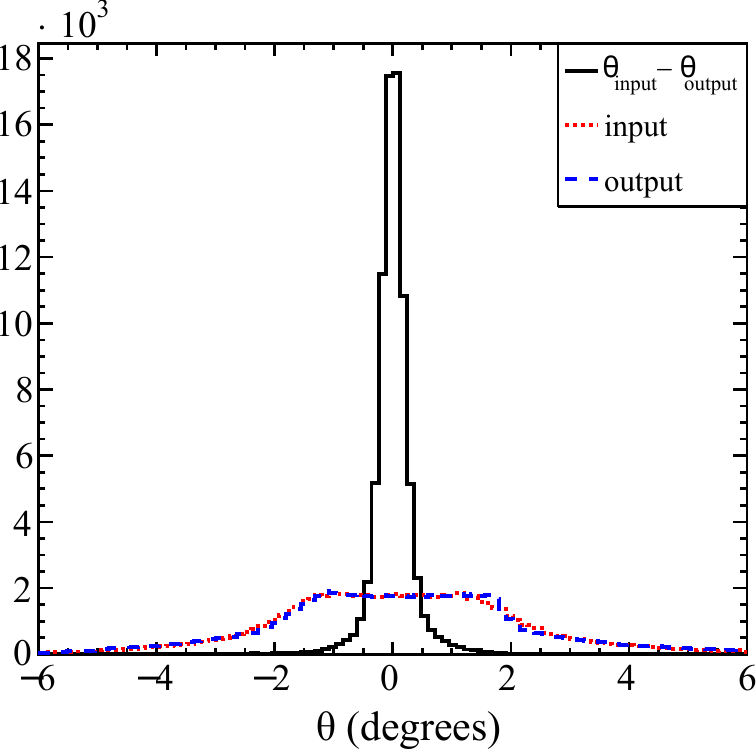}
  \end{center}
  \caption{\label{fig:nnC}The performance of the network correction to 
    calculate the physics scattering angle  $\theta$ (in radians). The red 
  histogram shows the simulation input data used to train the network. The blue 
histogram shows the trained network result and the black histogram shows the  
difference between the nominal (red, small dash) and network output (blue, long dash) results.}
\end{figure}

Using the hits in BETA and knowledge of the target's 5\,T field, the trajectory 
of the scattered electron was reconstructed to allow the determination the 
kinematics of each event. While na\"ive, straight-line tracks from $x$ and $y$ 
calorimeter hits to the target gave initial physics scattering angles $\theta$ 
and $\phi$, corrections were made to take into account the angle of incidence 
in the calorimeter and, more importantly, the bending of the electron in the 
magnetic field. The electron and positron $x$-$y$ position correction neutral network (b) was very similar to the network for photons,  shown in 
Figure~\ref{fig:nn}.
The final neural network (c) was 
trained to produce the physics scattering angles $\theta$ and $\phi$. Figure~\ref{fig:nnC} shows the network performance for the physics scattering angle 
$\theta$.

\subsection{Cherenkov Calibration}

\begin{figure}[htb]
	\begin{center}
		\includegraphics[width=0.90\columnwidth,trim=0mm 0mm 0mm 0mm,clip]{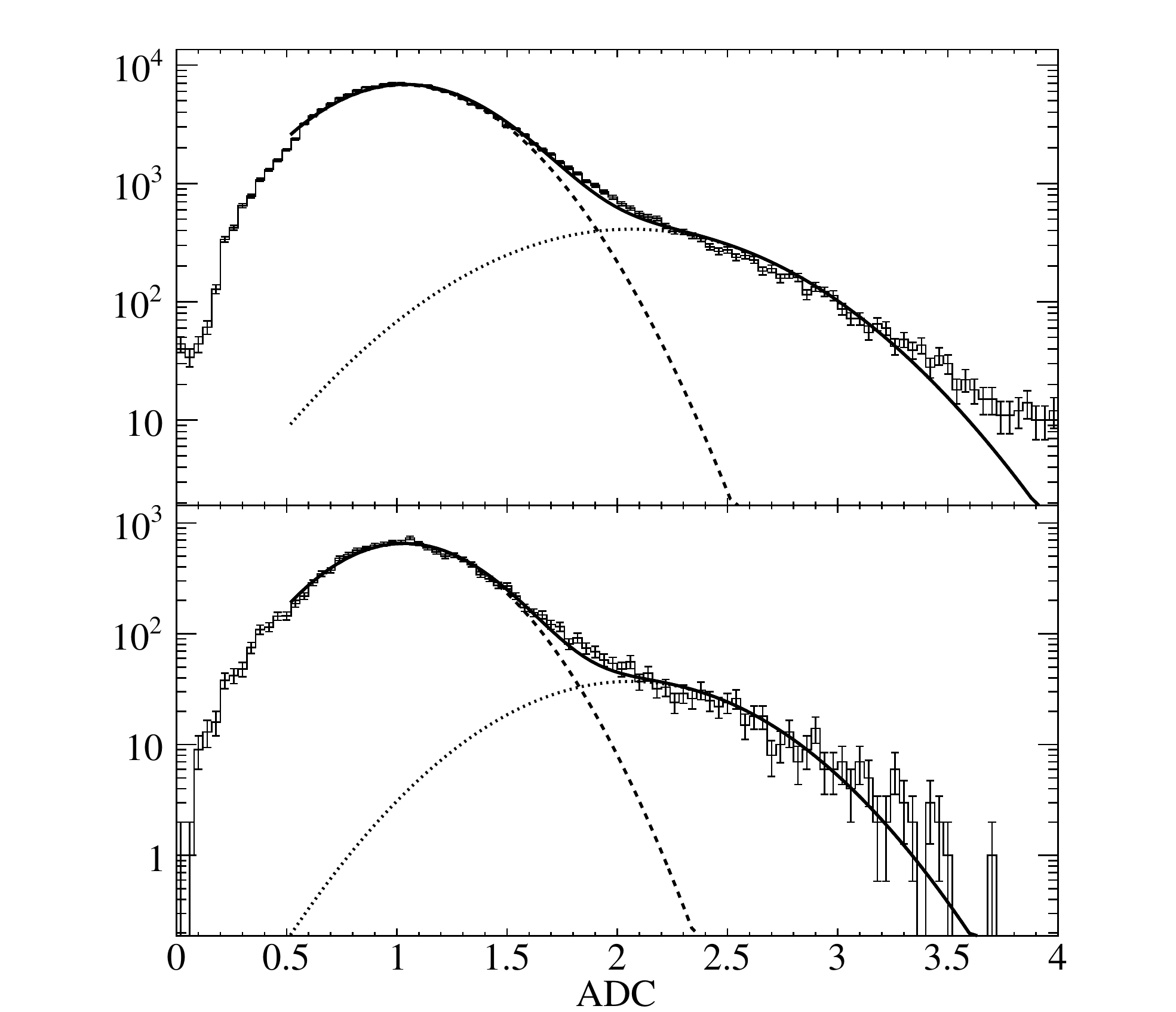}
	\end{center}
	\caption{\label{fig:cherenkovADCs}Cherenkov counter ADC spectrum for all the toroidal mirrors (top)
		and spherical mirrors (bottom).}
\end{figure}

Each of the Cherenkov's eight ADC spectra were normalized to their average 
single-electron track signal, which corresponded to roughly 18 photoelectrons.  
This provided an ADC spectrum calibrated to the number of electrons and 
positrons, as seen in Figure~\ref{fig:cherenkovADCs}, which shows a fit 
for the relative contribution of single and double tracks. These ``double tracks'' 
are electron--positron pairs produced outside the target field---either in the 
scattering window, front hodoscope, or Cherenkov window--- that travel co-linearly
after production to create a single cluster in the calorimeter. Pairs produced in the target separate due to the field, to be rejected as two-cluster events if both arrived 
in the calorimeter, or remain as background if only one arrived in the calorimeter (see section~\ref{sec:pairsym}). 
The single and 
double track signal fit results were used to estimate the double track 
background in an ADC window cut (see section~\ref{sec:event}).

\section{Asymmetry Analysis}

Because BETA was a new detector configuration, we discuss here the analysis 
framework required for its inclusive spin asymmetry measurements, leaving HMS 
analysis details to other works~\cite{liyanage,kang2}. Deep-inelastic scattering electron events 
detected in BETA were reconstructed, separated into kinematic bins, formed into 
yields based on the beam helicity, and corrected to produce physics asymmetries 
at each target field angle. These asymmetries take the form
\begin{linenomath}
\begin{equation}
A = \frac{1}{fP_BP_T}\frac{N_+-N_-}{N_++N_-},
\label{eq:asym}
\end{equation}
\end{linenomath}
for dilution factor $f$, beam and target polarizations $P_B$ and $P_T$, and 
corrected electron yields for each beam helicity $N_\pm$.  
Here the target and beam polarizations are applied as a single, charge averaged 
value for all events in each experimental run, while the dilution factor and 
the yields are functions of the kinematics of each event.

%
%
%
 
\subsection{Event Selection}
\label{sec:event}
\begin{figure}[h]
  \centering
  \includegraphics[trim=0mm 4mm 0mm 4mm,clip, width=0.5\textwidth]{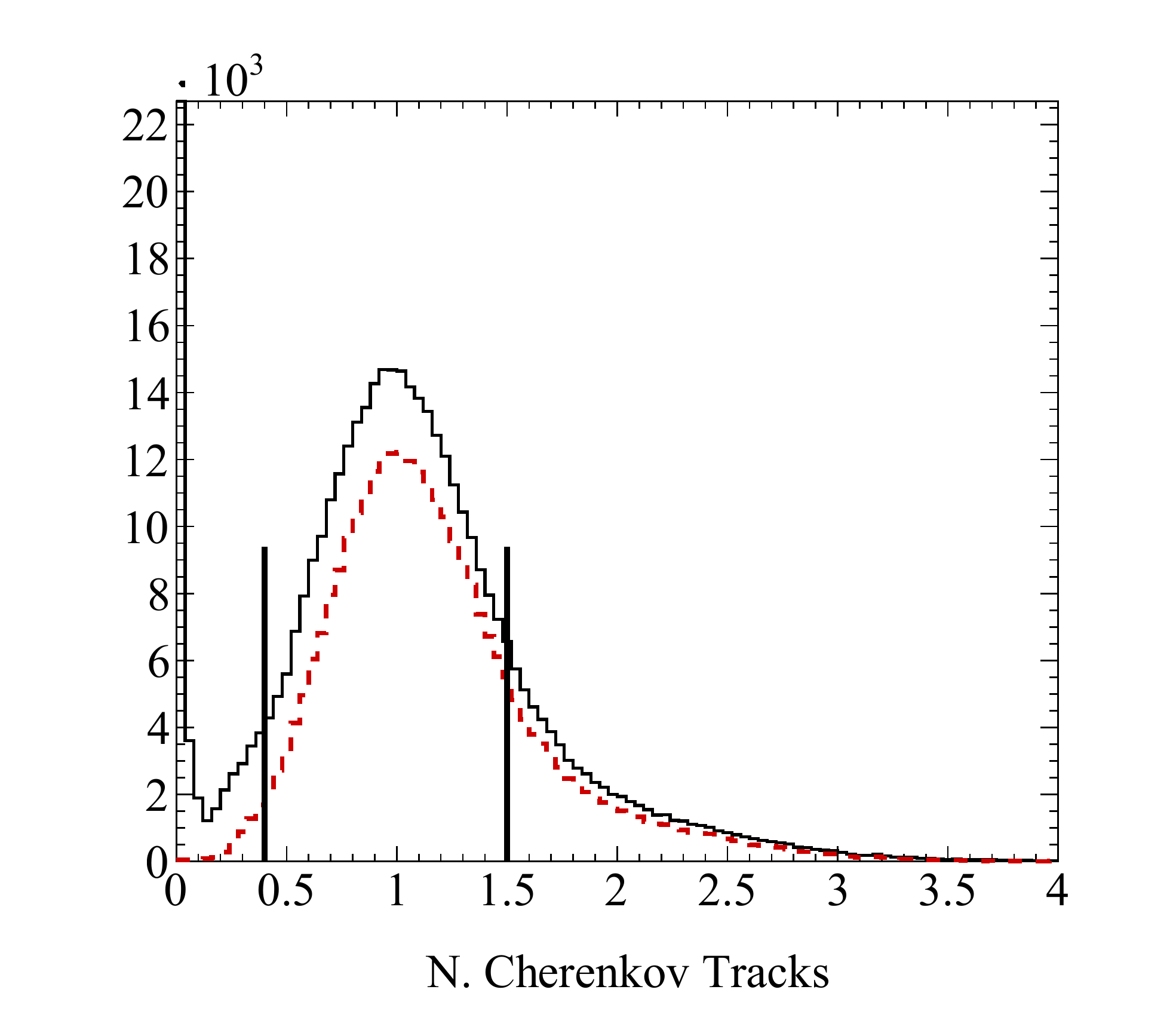}
  \caption{\label{fig:cerADCpeak}The Cherenkov ADC spectrum without (solid) and 
    with (dotted) a TDC cut. The Cherenkov ADC window cut is defined by the vertical 
  lines.}
\end{figure}

To minimize backgrounds and ensure that good electron events were counted in the 
yields, events were rejected if they did not meet the following criteria.  For 
asymmetry yields, only single cluster events in BigCal with a corresponding 
Cherenkov hit were taken. A cut was placed on the Cherenkov hit geometry, 
ensuring that the position in the calorimeter matched a hit in the correct 
Cherenkov sector. 
To reduce the systematic error due to the $\pi^0$ background subtraction 
(described in section \ref{sec:pairsym}), 
single clusters in BigCal below an energy cut of 900\,MeV were excluded.
The Cherenkov window cut provided a clean selection of single-track events and 
removed most of the background contribution from double-track events. The 
dominant source of double-track events came from pair production outside of the 
strong target magnetic field. The Cherenkov ADC window cut is shown in 
Figure~\ref{fig:cerADCpeak}.

\subsection{Asymmetry Measurements}

To extract physics spin asymmetries, SANE directly measured double-spin 
asymmetries with the target's magnetic field anti-parallel and at 80\degrees~to 
the beam. Reconstructed electron event yields from each helicity $n_\pm$ were 
used to form raw asymmetries $A_{180\degrees}$ and  $A_{80\degrees}$, as a 
function of their $x$ and $Q^2$ kinematic bins:
\begin{linenomath} \begin{equation}
A_{\textrm{raw}}(x,Q^2) = \frac{n_+(x,Q^2)-n_-(x,Q^2)}{n_+(x,Q^2)+n_-(x,Q^2)}.
\end{equation}
\end{linenomath}
These raw asymmetries must be first corrected for the effects of dead time in 
the data acquisition system, unequal total electron events in each helicity, 
and the dilution of the target by material other than the protons of interest.

\subsubsection{Charge Normalization and Live Time Correction}

Although the 30\,Hz, pseudo-random helicity flips of the beam produced nearly 
equal number of positive and negative helicity incident electrons, any 
imbalance in the beam charge between the two helicity states would introduce a 
false asymmetry. This effect was corrected by normalizing the asymmetry using 
total charge accumulated $Q_+$ and $Q_-$ from each helicity. The beam charge 
was measured by a cylindrical cavity which resonates at the same frequency as 
the accelerator RF in the transverse magnetic mode as the beam passes through 
the cavity.
The RF power of the resonance was converted by antennae in the cavity
into an analog voltage signal. This analog signal was processed into a
frequency which was then counted by scalers which were gated for beam helicity. 
A special set of data was taken to calibrate the beam current 
measured in the hall relative to the beam current measured by a Faraday
cup in the accelerator injector at various beam currents.
The scalers were injected into the datastream every two seconds, and 
experimental data was used only if the beam current was between 65 and 100~nA.

Typically, scalers measured the total number of accepted triggers, 
$n^{\textrm{acc}}_{\pm}$, and the total trigger events, 
$n^{\textrm{trig}}_\pm$, for each helicity.
To account for the computer livetime from either helicity due to event triggers 
that arrived while the data acquisition was busy, the corrected yield was divided 
by the computer livetime: $L_\pm = 
n^{\textrm{acc}}_{\pm}/n^{\textrm{trig}}_\pm$. Together, the charge 
normalization and livetime corrections resulted in corrected yields 
\begin{linenomath} \begin{equation}
N_\pm = \frac{n_\pm}{Q_\pm L_\pm},
\end{equation}\end{linenomath}
for raw counts $n_\pm$ of electron yields of each helicity, for each run, and 
as a function of kinematic bin.

\begin{figure}
  \begin{center}
    {%
    \includegraphics[width=0.9\columnwidth]{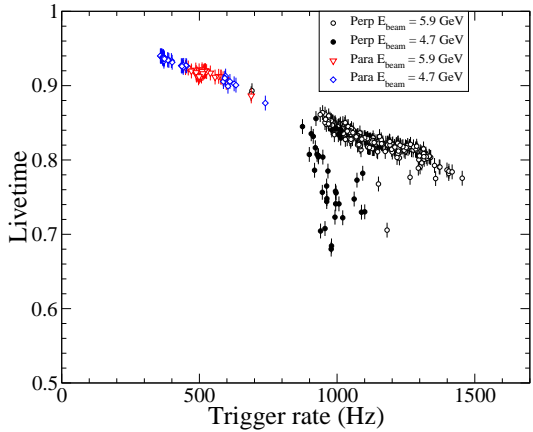}
}
  \end{center}
  \caption{The computer livetime for negative helicity events as a function of 
  negative helicity trigger rate.}
  \label{fig:livetime}
\end{figure}

Unfortunately, during SANE the total positive beam helicity trigger events from 
the scalers was not measured and therefore a direct measure of $L_{+}$ was not 
made. The total negative beam helicity trigger events were, however, recorded by the 
scalers, as were the accepted trigger events for both helicities.
The livetime for the negative helicity was calculated for each run from the scaler data.
Given the trigger rates of the experiment, the livetime
could be approximated  as $1  - \tau R^{\textrm{trig}}$, where $R^{\textrm{trig}}$ is the rate of
triggers and $\tau$ is the computer deadtime of the data acquisition system.
For each run, $\tau$ was determined from the negative helicity data and the livetime
for each helicity, $L_\pm$,  was calculated as $1  - \tau R^{\textrm{trig}}_\pm$.
A plot of the livetime for the negative helicity events for all the runs in the 
experiment is shown in Figure~\ref{fig:livetime}.
For most of the experimental data, the livetime measurement was consistent with 
$\tau \approx 160~\mu$sec.
However, the 4.7~GeV, perpendicular-target data shows  large
variations in the livetime with only small variation in trigger rate, implying that $\tau$ must have been fluctuating. The cause of this effect is 
not fully understood.

\begin{figure}
  \begin{center}
    {%
    \includegraphics[width=0.90\columnwidth]{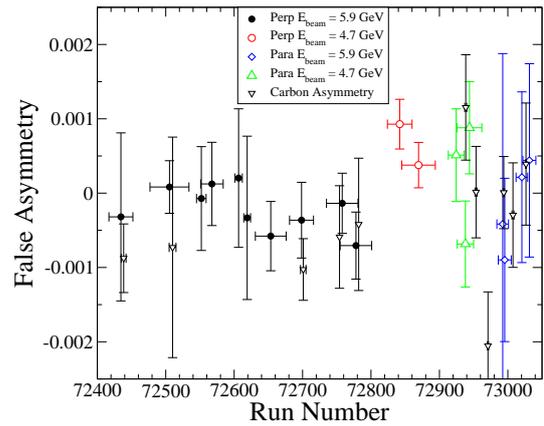}
}
  \end{center}
  \caption{The false asymmetry for pairs of run groups with opposite sign of 
  $P_B P_T$ versus run number.}
  \label{fig:false_asym}
\end{figure}

To check the effectiveness of the charge and livetime corrections to the data, 
a measurement of the false asymmetry was done using the trigger asymmetry, 
$A_{p,n}$, as measured with positive ($p$) or negative ($n$) combinations of 
beam, $P_B$, and target, $P_T$ polarizations. The false asymmetry was calculated as
\begin{linenomath} \begin{equation}
A_{\textrm{false}} = \frac{C_{p}A_{n}-C_{n}A_{p}}{C_{p}-C_{n}},
\end{equation}\end{linenomath}
and $C = P_B P_T$, with the $p (n)$ indicating the sign of $C$. 
In Figure~\ref{fig:false_asym}, the false asymmetry is plotted as
a function of run number.

\subsubsection{Packing Fraction} 

The ammonia target samples consisted of irregular beads roughly 2\,mm in 
diameter, cooled in a liquid helium bath and held with aluminum foil windows.  
Each sample differed slightly in the amount, size and shape of the beads used. To 
determine what portion of the target cell was ammonia, called the packing 
fraction $p_f$, experimental yields from the HMS were compared to simulation. 
A carbon disk target was utilized in specialized runs throughout the experiment 
to provide yields with a well-known cross section and density, giving
a normalization for the HMS acceptance and beam charge.
The electron yield was a linear function of the packing fraction $Y(p_f) = m p_f 
+b$, where $m$ and $b$ depend on the beam current, acceptance, partial 
densities and cross sections.

\begin{figure}
  \centering
      \includegraphics[width=0.9\columnwidth,trim=0mm 0mm 0mm 0mm, 
      clip]{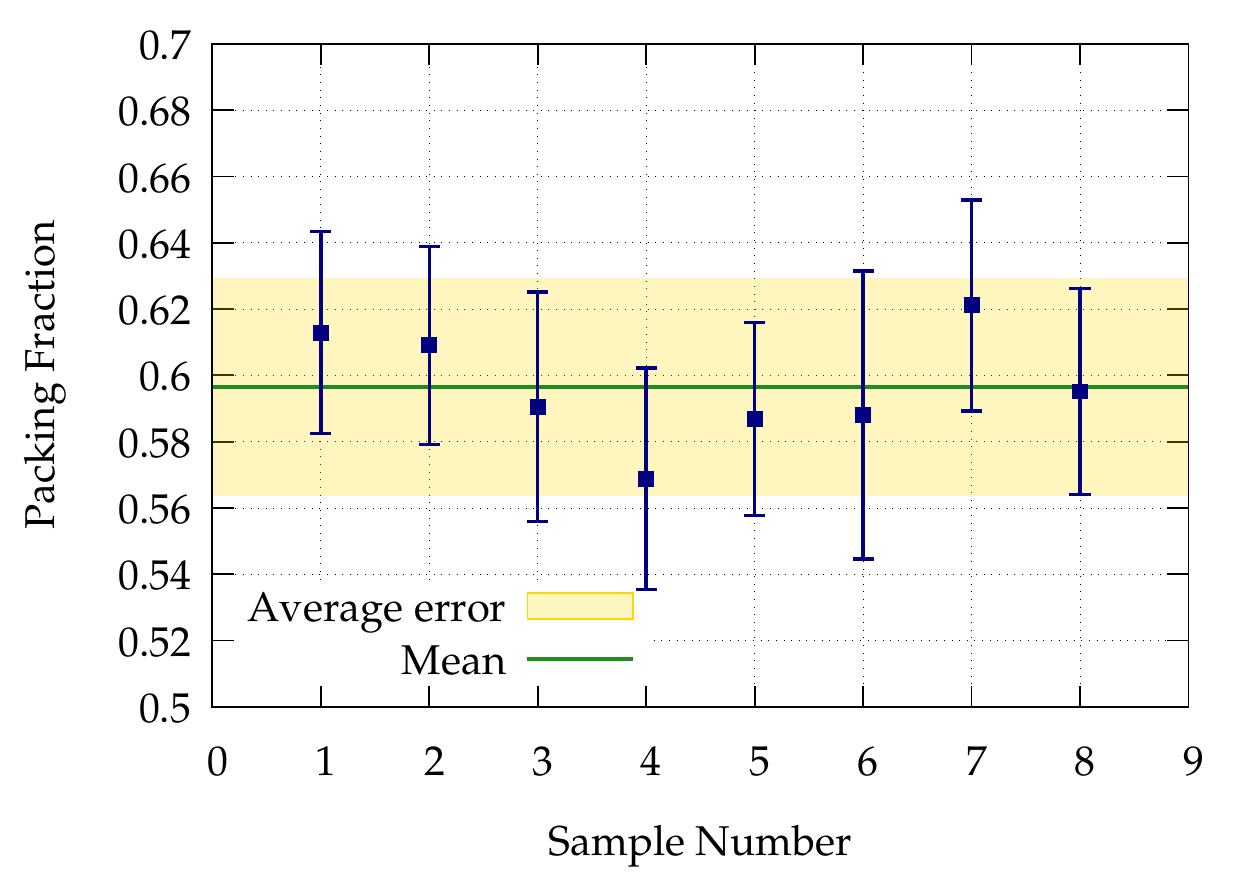}
  \caption{\label{fig:pf} Packing fractions for all target material samples 
  used during SANE, showing averaged value and error.}
\end{figure}

Using this linear relation, the packing fraction of a given 
sample was determined by interpolating between two reference points on the line, as simulated 
from a Monte Carlo.  The Hall C HMS single 
arm Monte Carlo---based on an empirical fit of inelastic cross 
section~\cite{pbec_F1F21,pbec_F1F22} and containing realistic HMS, target and 
field geometries---was run with target packing fraction set to 50\%, and again with packing fraction set to 60\%.
The simulated yields from these two points of known packing fraction provided the necessary line for interpolating the
target sample's packing fraction from the given HMS experimental yields. 
Figure~\ref{fig:pf} shows the calculated packing fractions for all SANE target 
material samples.

\subsubsection{Dilution Factor}
\begin{figure}[h]
  \centering
      \includegraphics[width=0.85\columnwidth,trim=10mm 5mm 10mm 10mm, 
      clip]{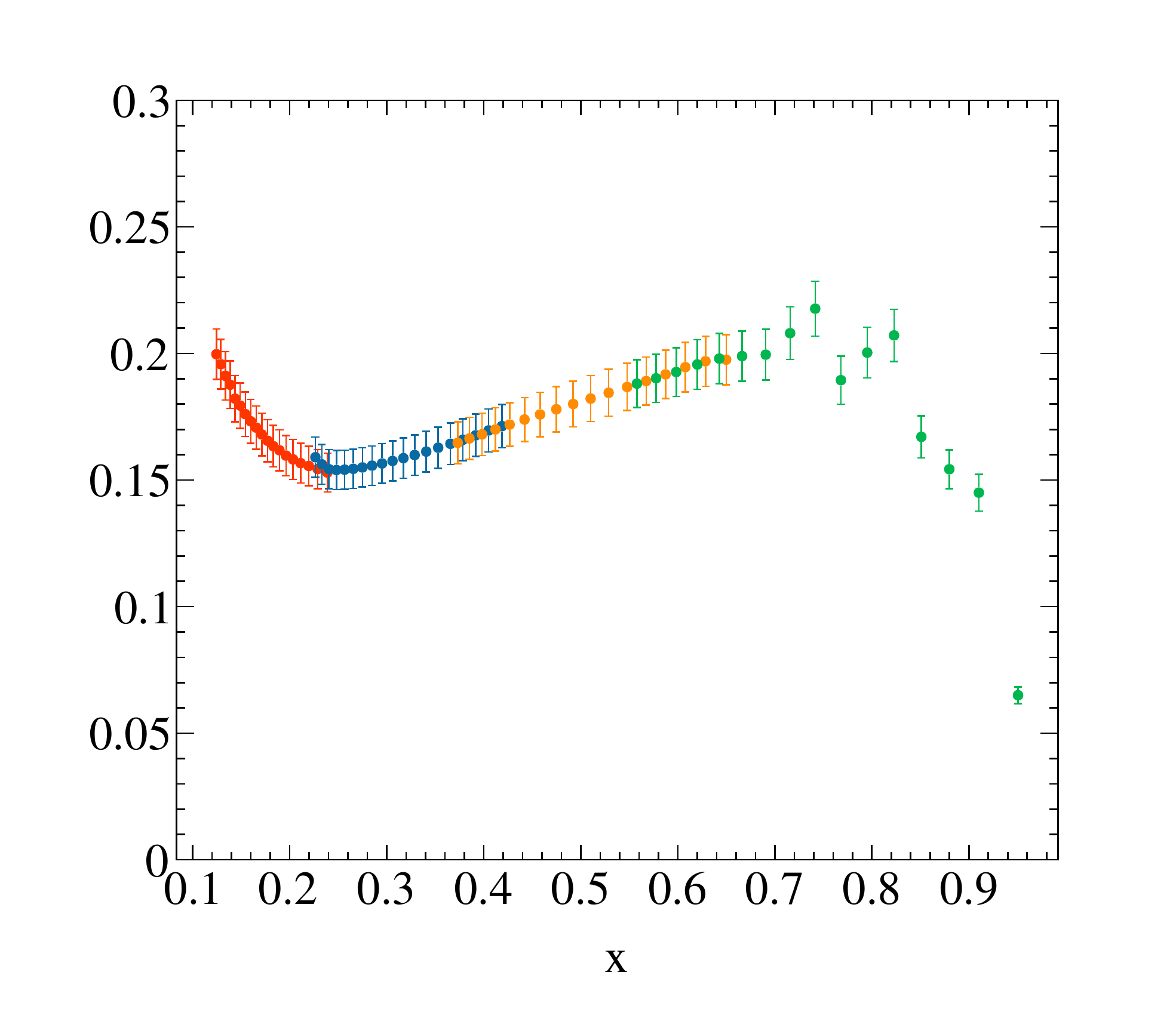}
  \caption{\label{fig:df}
    The dilution factor calculated for run 72925 as a 
    function of x, showing the increasing contribution from the elastic tails at 
    lower energies (i.e.  lower x). Each color represents a different $Q^2$ bin.
  }
\end{figure}

The dilution factor, $f$, is a kinematics dependent correction to the measured 
asymmetries to account for contributions of unpolarized nucleons in the target. 
Essentially a ratio of the cross-sections of the polarized protons to the 
nucleons of all other materials in the target cell, the dilution factor was 
calculated for each experimental run as
\begin{linenomath} \begin{equation}
f(W,Q^2) =\frac{N_1\sigma_1}{N_1\sigma_1+N_{14}\sigma_{14}+ \Sigma N_A\sigma_A},
\label{eq:df}
\end{equation}\end{linenomath}
for number densities $N_A$ of each nuclear species present in the target of 
atomic mass number $A$, and radiated, polarized cross-sections 
$\sigma_A(W,Q^2)$~\cite{rondon2005packing}. This factor covers not only the 
protons (1) and nitrogen (14) in the ammonia sample, but must also include 
other materials such as helium (4) and aluminum (27). Substituting numeric 
values for this specific target, the dilution factor is expressed in terms of 
these cross sections and the packing fraction $p_f$ as
\begin{linenomath} \begin{equation}
f =  \left(1 +\frac{\sigma_{14}}{3\sigma_1}+
0.710 \left[\frac{4}{3 p_f}-1\right]\frac{\sigma_4}{3\sigma_1}+
\frac{0.022}{p_f}\frac{\sigma_{27}}{3\sigma_1}\right)^{-1}.
\label{eq:df2}
\end{equation}\end{linenomath}
Cross sections for each species needed for Equation~\ref{eq:df2} were 
calculated from empirical fits to structure functions and form factors, and 
included all radiative corrections used later in the analysis. The dilution 
factor for a typical run is shown in Figure~\ref{fig:df} in $x$ bins.

\subsubsection{Target Radiation Thicknesses}

The thickness of each radiator in the scattering chamber was required for the 
calculation of external radiative corrections. Table \ref{tab:thick} shows the 
radiation thickness for all materials traversed by the beam passing through the 
target, for a nominal packing fraction of 0.6, as well as the percentage of 
radiation length $\chi_0$.
\begin{table}[h]
  \begin{center}
    \begin{tabular}{llL{1cm}c}
      \toprule
      Component		& Material  & Thickness (mg/cm$^2$) & $\chi_0$ (\%) \\
      \midrule
      Target Material & $^{14}$NH$_3$ &  1561 & 3.82\\ 
      Target Cryogen & LHe &  174 & 0.18\\
      Target Coil & Cu &  13 & 0.10\\
      Cell Lid & Al &  10 & 0.04\\
      Tail Window & Al &  27 & 0.12\\
      Rad Shield& Al &  7 & 0.03\\
      N Shield& Al &  10 & 0.04\\
      Beam Exit& Be & 24 & 0.04\\
      \multirow{2}{*}{Vacuum Windows} & Be &  94 & 0.14\\ 
                                      & Al & 139 & 0.58\\ 
      \cmidrule{2-4}
      &\multicolumn{2}{l}{80\degrees\ Total, Before Center}  &2.98\\
      &\multicolumn{2}{l}{80\degrees\ Total, After Center}  &2.36\\
      &\multicolumn{2}{l}{180\degrees\ Total, Before Center}  &2.54\\
      &\multicolumn{2}{l}{180\degrees\ Total, After Center}  &2.36\\

      \bottomrule
    \end{tabular}
    \caption[Table of component thicknesses for radiative corrections]{Table of 
      target component thicknesses for radiative corrections.  Total thicknesses before 
      and after the center of the target are given for each magnet orientation 
    configuration.}
    \label{tab:thick}
  \end{center}
\end{table}

\subsubsection{Polarized Nitrogen Correction}

While the dilution factor correction accounts for scattering from material 
other than protons, it does not take into account the effect of any 
polarization of such material in the asymmetry. Nitrogen, in particular, 
provides a third of the polarizable nucleons in ammonia. During usual  DNP 
conditions, the polarization of the  spin-1/2 protons  ($P_p$) and spin-1 nitrogen 
($P_N$) in $^{14}$NH$_3$ are related as 
\begin{linenomath} \begin{equation}
P_N = \frac{4\tanh((\omega_N/\omega_p)\arctanh(P_p))}{3+\tanh^2((\omega_N/\omega_p)\arctanh(P_p))},
\end{equation}\end{linenomath}
where $\omega_N$ and $\omega_p$ are the  $^{14}$N and proton Larmor 
frequencies~\cite{Adeva}. At maximum proton polarizations of 95\%, the nitrogen 
polarization will be only 17\%.
In addition, in nitrogen a nucleon's spin is aligned anti-parallel to the spin of 
the nucleus one third of the time~\cite{orpolcorr}. These effects together 
result in a maximum polarization of anti-parallel nitrogen nucleons of roughly 
2\%, which results in an added systematic error to the asymmetries of less than 
half a percent.

\subsubsection{Pair-symmetric background subtraction}
\label{sec:pairsym}
At lower scattered electron energies, the pair-symmetric background becomes 
significant, and  pair conversions that happen in, 
or very near, the target cannot be completed rejected.
Cherenkov window cut (shown in Figure 
\ref{fig:cerADCpeak}) was only capable of removing double-track events---tracks which 
produce twice the amount of Cherenkov light as a single electron track. 
Double-track events are the result of $e^+$--$e^-$ pairs which are produced outside of the 
target. These are not significantly deflected by the magnetic field, and thus
appear as one cluster with twice the expected Cherenkov light, easily
removed by the Cherenkov window cut. However, pairs produced in the target 
material are significantly deflected, causing only one particle to be detected 
in BETA. These events cannot be removed with selection cuts and are 
misidentified as DIS electrons.

To compensate for the pair-symmetric background, the scattering asymmetry $A$ 
from Equation~\ref{eq:asym} was corrected with 
\begin{linenomath}\begin{equation}
A_\text{corrected}=A/f_\textrm{BG} - C_\textrm{BG}.
\end{equation}\end{linenomath}
where $f_\textrm{BG}$ is the background dilution, and $C_\textrm{BG}$ is the 
pair-symmetric background contamination of the measured asymmetry.
The background dilution term corrects for the unpolarized background 
contribution to the total yield, and the contamination term removes any 
background asymmetry contributing to the measured asymmetry.


The dominant source of pair-symmetric background events came from conversion
of $\pi^0 \rightarrow \gamma \gamma$ decay photons. Events 
passing the selection cuts were either inclusive electron scattering events or 
pair-symmetric background events. The background dilution is then
 $f_\textrm{BG} = 1-f_\textrm{SANE}$, 
where $f_\textrm{SANE}=n_\textrm{BG}/n_\textrm{total}$ is the ratio of background 
to total scattering events.
The contamination term is defined as 
\begin{equation}
  C_\textrm{BG}=\frac{f_{\pi^0}^p}{f}\frac{A_{\pi^0}f_\textrm{SANE}}{1-f_\textrm{SANE}}\,,
\end{equation}
where $A_{\pi^0}$ is the inclusive $\pi^0$ asymmetry, and $f_{\pi^0}^p/f$ is the 
ratio of target dilution factors for $\pi^0$ production and electron 
scattering. The target dilution for electron scattering is defined in 
Equation~\ref{eq:df}, and the background target dilution, $f_{\pi^0}^p$, is similarly 
defined using cross sections for inclusive $\pi^0$ production. This ratio can 
be roughly approximated as unity ($f_{\pi^0}^p/f \simeq 1$) as it is well 
within the systematic uncertainties.

\begin{figure}[h]
  \centering
      \includegraphics[width=0.8\columnwidth,trim=10mm 5mm 10mm 10mm, 
      clip]{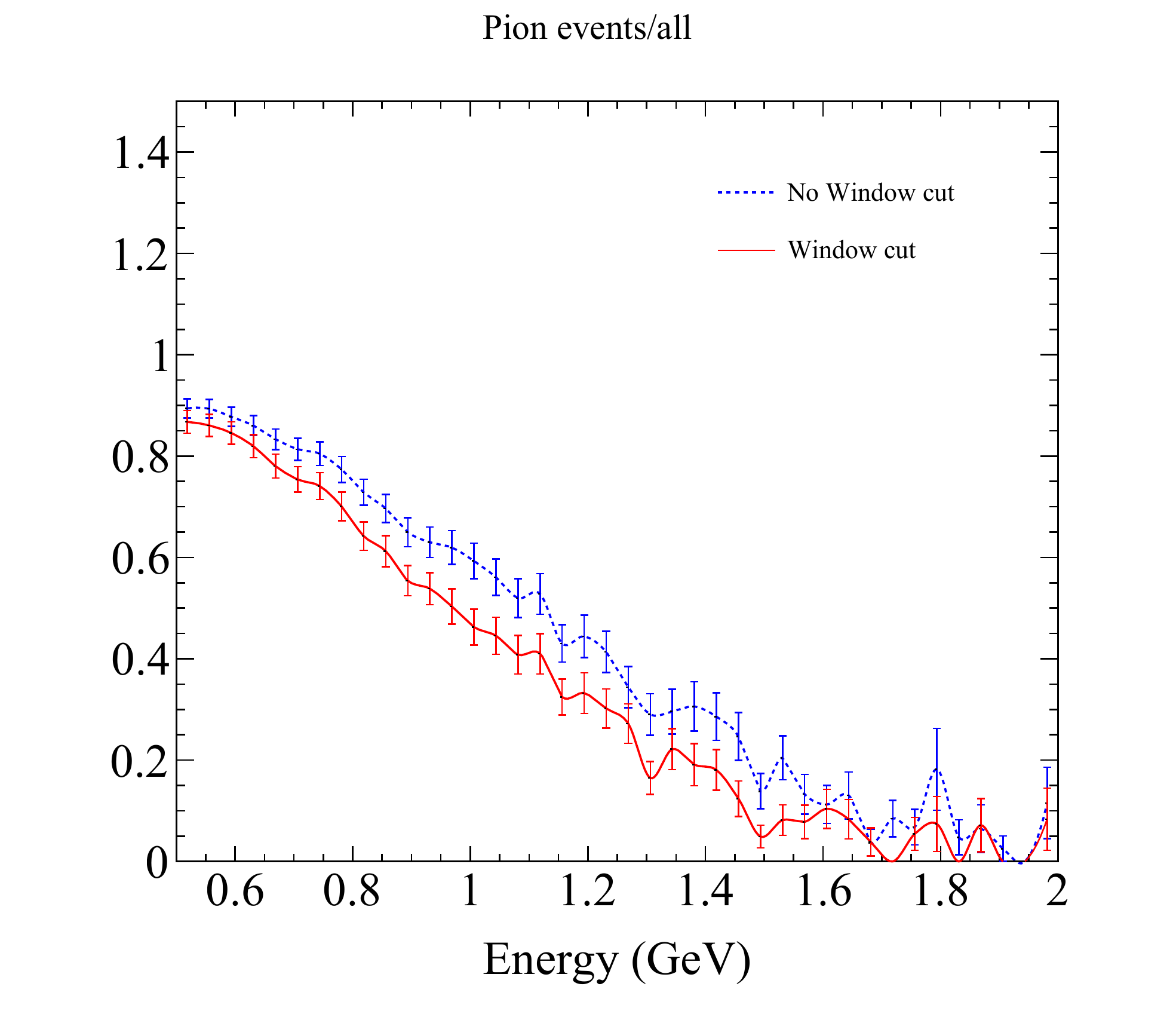}
  \caption{\label{fig:pairBGratio}Simulations results for the pair symmetric 
    background ratio $f_\textrm{SANE}$ as a function of the 
    scattered electron energy.  The lower curve is the ratio with the 
    Cherenkov ADC window which removes the background contributions from 
  pairs converted in material outside of the target cell.  }
\end{figure}

Simulations of the $\pi^0$ background and inclusive electron scattering were 
employed to determine $f_\textrm{SANE}$ which is shown in 
Figure~\ref{fig:pairBGratio}.  A FORTRAN routine to model inclusive pion 
production by J.\ O'Connell \cite{OConnell} was updated using photoproduction 
cross section data from the Yerevan Physics Institute \cite{Alanakyan} to 
improve the cross section reproduction to better than 15\% in the kinematics of 
interest.
The updated pion production model also displayed good agreement when
compared to charged pion electroproduction data \cite{Heimlich}.
The asymmetry of the pair-symmetric background, $A_{\pi^0}$, was estimated from 
fits to charged pion, parallel and transverse, asymmetry data taken on
polarized $^{15}$NH${_3}$ in SLAC experiments E143 and E155x.
Data for both pion charges were averaged as a substitute for
$\pi^0$.
See \ref{sec:pion_ap} for a further discussion of the pion asymmetries.

\subsection{Beam and target systematic errors}

Table~\ref{tab:error} shows an overview of SANE systematic error contributions 
from the beam and target systems, which enter Equation~\ref{eq:asym} as 
kinematics independent normalizations, and the kinematics dependent dilution 
factor. The error in the target polarization was the single largest 
contribution, and stems from the NMR polarization measurements. The NMR can be 
affected by minute shifts in the material beads over time and topological 
differences in dose accumulation around the coils embedded in the material. The 
thermal equilibrium measurements on which the enhanced NMR signals were 
calibrated also add error, with the temperature measurement of the material 
contributing significantly. Looking at the differences in the TE measurements 
over the experimental life of any given material gives an indication of the 
error. For example, material four's 3 TE measurements had a standard deviation 
of 8\% around their mean, while material five had the same number of TE's with 
a 2\% standard deviation. A detailed discussion of error in DNP targets from 
the SMC collaboration can be found in reference~\cite{Adams}.

The global error in the beam polarization measurements contributes 1\%, while the 
fit used to apply the measurements at varied beam energies will add another 
half percent. The dilution factor's uncertainty is based on statistical error 
in the measurement of the packing fraction and from the simulation.

\begin{table}[ht]
  \begin{center}
    \begin{tabular}{lc}
      \toprule
      Source &  Error on Asymmetry   \\ \midrule 
      Beam polarization  & 1.5\%  \\ 
      Target polarization & 5.0\%  \\ 
      Nitrogen correction  & 0.4\%\\
      Dilution factor  & 2.0\%\\
      \cmidrule{2-2}
      Combined  & 5.6\% \\
      \bottomrule
    \end{tabular}
    \caption{Table showing systematic errors from the polarized beam and target.}
    \label{tab:error}
  \end{center}
\end{table}

\section{Conclusion}

\begin{figure}[h]
  \centering
      \includegraphics[width=0.8\columnwidth,trim=2mm 5mm 10mm 10mm, 
      clip]{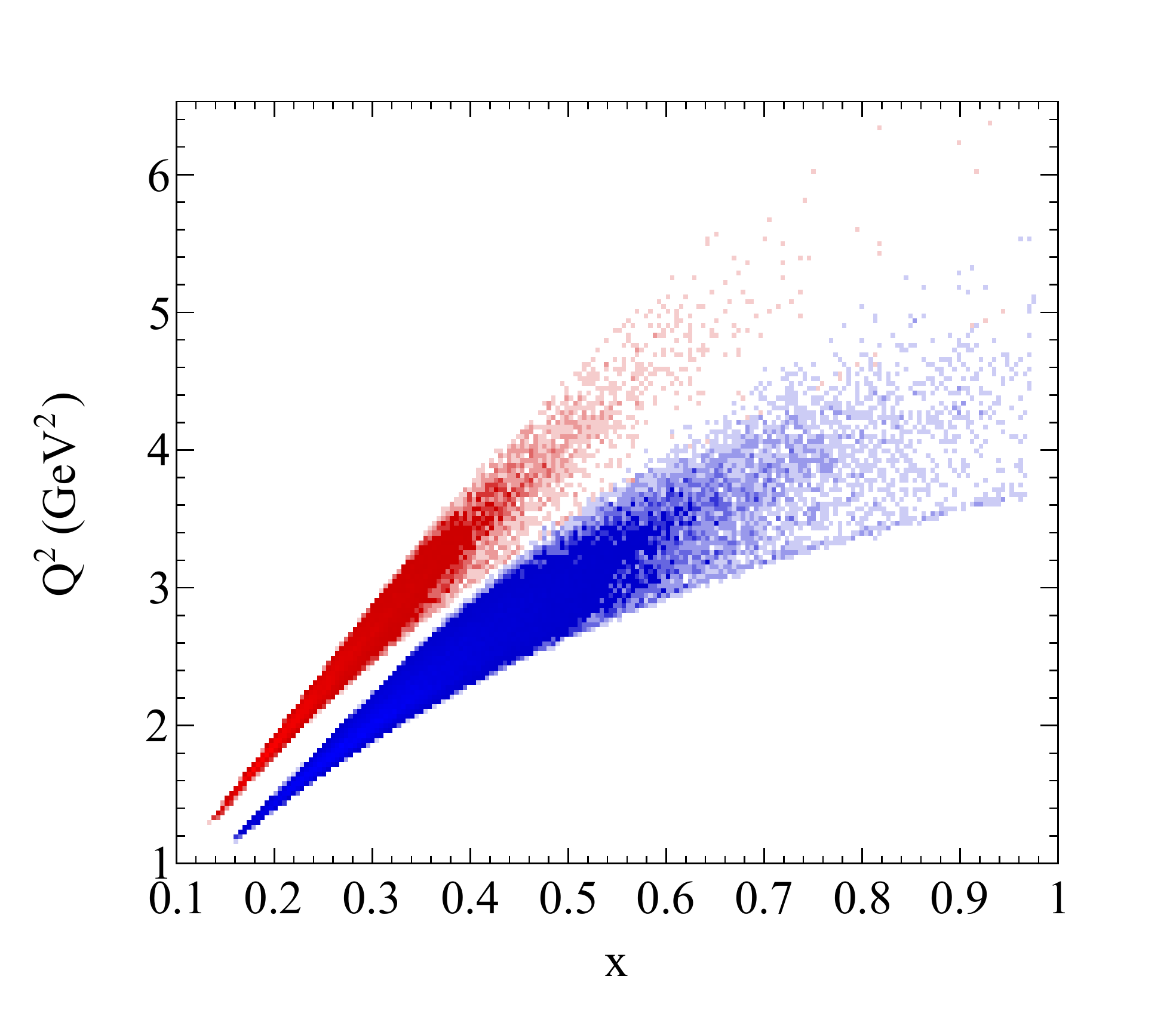}
  \includegraphics[width=0.8\columnwidth,trim=2mm 5mm 10mm 10mm, 
  clip]{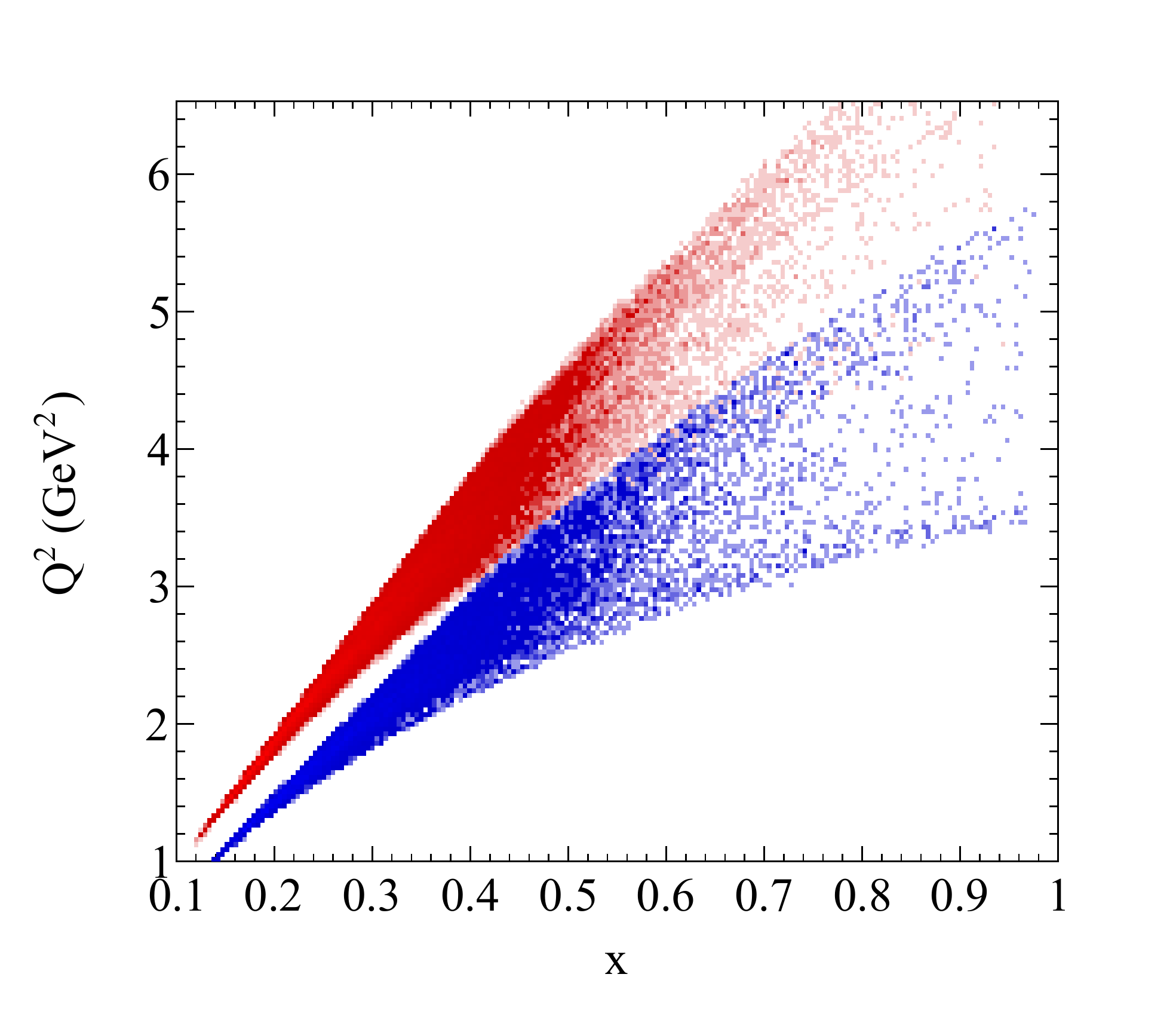}
  \caption{\label{fig:kine}The kinematic coverage of SANE events, before cuts, 
    with target oriented parallel (top) and at 80\degrees~to the beam (bottom). 
    Red points represent 5.9\,GeV beam energy coverage, while blue points show 
  4.8\,GeV. }
\end{figure}

Through a combination of a novel, wide-acceptance electron arm, and a 
rotatable, solid polarized proton target, the Spin Asymmetries of the Nucleon 
Experiment has significantly expanded the world's inclusive spin structure data 
for the proton. By taking spin asymmetry measurements with the target oriented 
at parallel and near perpendicular, model-independent access to virtual Compton 
asymmetries $A_1^p$ and $A_2^p$ on the proton was possible with the only input 
being the well measured ratio of longitudinal to transverse unpolarized cross 
sections $R_p$. The only other sources of model independent proton $A_1$ 
measured in the same experiment are SLAC's E143 at 29 GeV~\cite{e143} and E155 
at 48 GeV~\cite{e155}, and the JLab's RSS~\cite{RSS}.  SANE's kinematic 
coverage (shown in Figure~\ref{fig:kine}) represents a crucial improvement to the 
world's data of inclusive proton scattering, particularly with a perpendicular 
target, filling in gaps in $x$ coverage to allow integration for moments of 
structure functions, such as $d_2$. Forthcoming letters will present the 
physics results of these efforts.

\section*{Acknowledgements}

We would like to express our sincerest gratitude to the staff and technicians 
of Jefferson Lab for their indispensable support during the running of SANE.  
We especially thank the Hall~C and Target Group personnel, who saw a 
technically challenging experiment through significant hardship to a successful 
end. This material is based upon work supported by the U.S. Department of Energy, Office of Science, Office of Nuclear Physics under contract DE-AC05-06OR23177.
This work was also supported by DOE grants DE-FG02-94ER4084 and DE-FG02-96ER40950.

\appendix
\section{Inclusive pion asymmetries}
\label{sec:pion_ap}
\begin{figure}	
	\begin{center}
		\includegraphics[width=0.98\columnwidth]{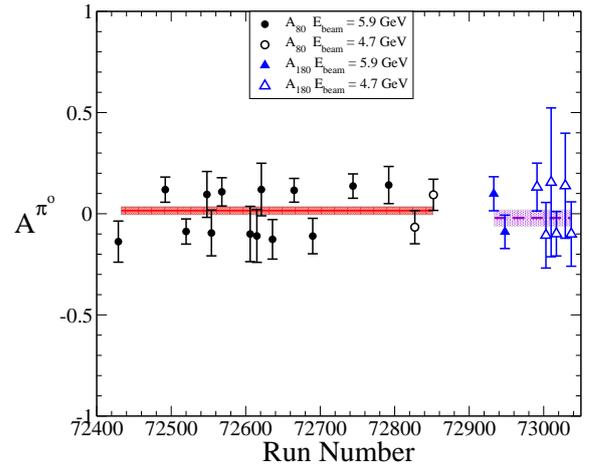}
	\end{center}
	\caption{The inclusive $\pi^{0}$ production  spin asymmetry, $A^{\pi^0}$,  plotted versus experiment's run number
	for anti-parallel,  $A_{180}$, and nearly perpendicular, $A_{80}$, target field directions for both beam energies.
    Weighted averages and error bands for 180\degrees (80\degrees) asymmetries are shown as red solid (violet dashed) line and shaded box.}
	\label{fig:SANE_apizero}
\end{figure}

The SANE experiment directly measured the $\pi^{0}$ spin asymmetries in both field directions
and at both beam energies~\cite{Ndukum}. The event selection criterion for $\pi^{0}$ events was
 two clusters in the calorimeter with a minimum separating distance of 20~cm,
  each cluster having greater than 0.6~GeV energy, and no signal in the Cherenkov detector.
The  $\pi^{0}$ energy ranged from 1.2 to 2.75~GeV.
 With the limited statistics, spin asymmetries were calculated by integrating 
the entire kinematic coverage in angle and energy. In Fig~\ref{fig:SANE_apizero}, the
 $\pi^{0}$ spin asymmetries are plotted as a function of the experiment's run number
 for both beam energies and field directions.
 Combining data from both beam energies, the weighted average of the nearly perpendicular  ($A_{80}$) and anti-parallel  ($A_{180}$) asymmetries 
 are 0.015 $\pm$ 0.019 and  -0.020 $\pm$ 0.040, respectively. 
The weighted averages are plotted in Figure~\ref{fig:SANE_apizero}  
 as a red solid (a violet dashed)  line with the error band shown by the shaded box for  $A_{180}$ ($A_{80}$).

Given the limited statistics of the SANE measurement for the inclusive pion asymmetry, data from previous
experiments was used to determine the inclusive pion asymmetry needed for background subtraction.
The spin structure experiments at SLAC (E143 \cite{e143}, E155 \cite{e155}, E155x \cite{e155x}) 
took  inclusive charged pion data as part of their systematic background studies. 
In addition, E155 took dedicated data on longitudinal hadron and pion asymmetries \cite{e155-3}.
The SLAC experiments measured spin asymmetries for target field directions that were parallel
and nearly perpendicular (at 92.4$^{\circ}$) to the beam directions. 
The data sets were taken from references \cite{Toole} and \cite{Benmouna:2001dm}.

The inclusive pion spin asymmetries can be parametrized as a function of the pion
transverse momentum, $P_T = p_\pi\sin(\theta_\pi)$, where $p_\pi$ and $\theta_\pi$ are 
the pion's outgoing momentum and angle.
The SLAC data is taken at larger pion momentum (between 10 to 30~GeV/c) and small 
forward angles (2.75$^{\circ}$ to  7$^{\circ}$) while the SANE
data is taken at smaller pion momentum (between 1.2 and 2.75~GeV/c) and larger angles (between
30$^{\circ}$ and 50$^{\circ}$).  Therefore, the SANE and SLAC experiments cover a comparable
range of P$_T$. The $\pi^{0}$ background for the SANE experiment has a lower limit of $P_T \approx 0.75$~GeV.

\begin{figure}	
	\begin{center}
		\includegraphics[width=0.98\columnwidth]{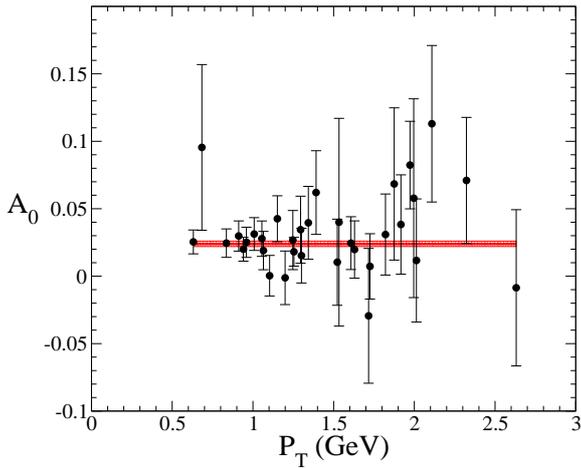}
	\end{center}
	\caption{SLAC pion production spin asymmetries for parallel target field direction  
	plotted as a function of $P_T$. A weighted average is shown as a red line, with the error band as a shaded box. }
\label{fig:SLAC_apara}
\end{figure}

The SLAC charged pion inclusive parallel spin asymmetries
are plotted as a function of $P_T$ in  Fig~\ref{fig:SLAC_apara}. 
The parallel data do not show any significant dependence on P$_T$ and the weighted average of the data has a 
$\chi$-squared per degree of freedom below one. The weighted average of the SLAC parallel asymmetry, $A_{0}$ data is 
0.024 $\pm$ 0.002 and is plotted as a solid red line in  Fig~\ref{fig:SLAC_apara} with the error band shown by the shaded box.
The SANE experiment used $^{14}$N in the ammonia target and SLAC used  $^{15}$N, so the SLAC asymmetry needs to
be multiplied by 14/15 to be compared with the SANE measurement. In addition, the parallel 
target field was at 180$^{\circ}$ 
for SANE compared to 0$^{\circ}$, so for SANE the asymmetry becomes -0.022 $\pm$ 0.002. The 
$\pi^{0}$ parallel asymmetry measure by SANE agrees with the SLAC measurement, but the SANE result has a 
much larger error bar. For the purpose of  $\pi^{0}$  background 
subtraction discussed in Sec.~\ref{sec:pairsym}, the SLAC weighted average was used.

\begin{figure}	
	\begin{center}
	\includegraphics[width=0.98\columnwidth]{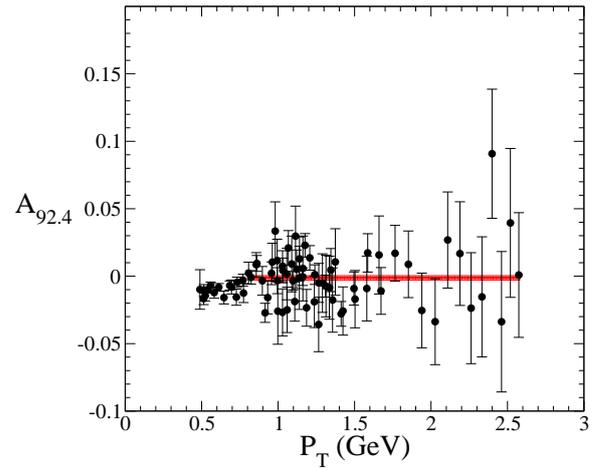}
    \end{center}
	\caption{SLAC pion production spin asymmetries for nearly perpendicular target field direction 
		  plotted as a function of $P_T$. A weighted average is shown as a red line, with the error band as a shaded box. }
	\label{fig:SLAC_aperp}
\end{figure}

The SLAC charged pion inclusive near perpendicularly  spin asymmetries
are plotted as a function of $P_T$ in  Fig~\ref{fig:SLAC_aperp}. 
The  data do not show any significant dependence on P$_T$ above 0.8~GeV/c and the weighted average of the data has a 
$\chi$-squared per degree of freedom below one. The weighted average of the SLAC data 
(corrected for $^{14}$N)
is $A_{92.4} = -0.0012 \pm 0.0016$ and is plotted as a solid red line 
in  Fig~\ref{fig:SLAC_aperp} with the error band shown by the shaded box.
 The perpendicular asymmetry at 90$^{\circ}$, $A_{90}$, is equal to $[A_{92.4}  - A_{0}\cos(92.4\degrees)]/\sin(92.4\degrees)$. 
Using the SLAC $A_{0}$ , then  $A_{90} = -0.0003 \pm 0.0016$. 
For the background subtraction discussed in Sec.~\ref{sec:pairsym}, $A_{90}$ was taken to be zero and the
error was applied part of overall systematic uncertainty.

\section*{References}
\bibliography{sane-nim}
\bibliographystyle{elsarticle-num}

\end{document}